\documentclass[12pt]{amsart}

\usepackage{amscd,amssymb,amsmath,latexsym,enumitem,blkarray}
\usepackage[mathscr]{euscript}
\usepackage{epsfig}

\numberwithin{equation}{section}

\usepackage{tikz}
\usetikzlibrary{arrows}

\usepackage{multicol}

\allowdisplaybreaks %allows align to break across pages

\usepackage[normalem]{ulem} %allows sout

\newcommand{\C}{\mathbb C}
\newcommand{\R}{\mathbb R}
\newcommand{\ol}{\overline}
\newcommand{\im}{\operatorname{Im}}

\newcommand{\Tr}{\operatorname{Tr}}
\newcommand{\re}{\operatorname{Re}}
\newcommand{\Span}{\operatorname{Span}}
\newcommand{\Range}{\operatorname{Range}}

\makeatletter
\def\blfootnote{\xdef\@thefnmark{}\@footnotetext}

\makeatother

\newtheorem{theorem}{Theorem}[section]
\newtheorem{lemma}[theorem]{Lemma}
\newtheorem{proposition}[theorem]{Proposition}
\newtheorem{corollary}[theorem]{Corollary}
\theoremstyle{definition}
\newtheorem{definition}[theorem]{Definition}
\newtheorem{example}[theorem]{Example}
\newtheorem{remark}[theorem]{Remark}

\begin{document}
\title{GKSL Generators and Digraphs: Computing Invariant States}
\author{George Androulakis and Alexander Wiedemann}\blfootnote{The article is part Wiedemann's PhD dissertation, prepared at the University of South Carolina under
	the supervision of Androulakis.}

\address{Department of Mathematics, University of South Carolina, 
Columbia, SC 29208}

\email{giorgis@math.sc.edu and akw@math.sc.edu}


\begin{abstract}
	In recent years, digraph induced generators of quantum dynamical semigroups have been introduced and studied, particularly in the context of unique relaxation and invariance. In this article we define the class of pair block diagonal generators, which allows for additional interaction coefficients but preserves the main structural properties. Namely, when the basis of the underlying Hilbert space is given by the eigenbasis of the Hamiltonian (for example the generic semigroups), then the action of the semigroup leaves invariant the diagonal and off-diagonal matrix spaces. In this case, we explicitly compute all invariant states of the semigroup.
	
	In order to define this class we provide a characterization of when the Gorini-Kossakowski-Sudarshan-Lindblad (GKSL) equation defines a proper generator when arbitrary Lindblad operators are allowed (in particular, they do not need to be traceless as demanded by the GKSL Theorem). Moreover, we consider the converse construction to show that every generator naturally gives rise to a digraph, and that under certain assumptions the properties of this digraph can be exploited to gain knowledge of both the number and the structure of the invariant states of the corresponding semigroup.
\end{abstract}


\maketitle
\section{Introduction}

\subsection{Exposition}

The Schr\"odinger picture time evolution of an open quantum system with finitely many degrees of freedom is, under certain limiting conditions, described in terms of a quantum dynamical semigroup (QDS) $(T_t)_{t\geq0}:M_N(\C)\to M_N(\C)$ (see e.g. \cite{A,AL}), where $M_N(\C)$ denotes the $N\times N$ matrices with complex entries. Each such QDS can be written as $T_t=e^{t\mathcal{L}}=\sum_{k=0}^\infty t^n\mathcal{L}^n/n!$ for some $\mathcal{L}$ called the generator of the QDS. Famously, simultaneous results of Gorini-Kossakowski-Sudarshan in \cite{GKS} and Lindblad in \cite{L} show that every QDS generator can be written as $\mathcal{L}(\rho)=-\imath[H,\rho]+\frac{1}{2}\sum c_{ij}([F_i,\rho F_j^\ast]+[F_i\rho,F_j^\ast])$, the now-called GKSL form (see Theorem~\ref{GKSL}). We call $H$ the Hamiltonian of the QDS.

Of particular interest are the digraph induced generators (where digraph means directed, positively weighted graph; see Section~\ref{digraphsection}), which we define as those of the form \begin{equation}\label{generic}\mathcal{L}(\rho)=-\imath[H,\rho]+\frac{1}{2}\sum_{i\neq j}\gamma_{ij}\left(\left[E_{ij},\rho E_{ij}^\ast\right]+\left[E_{ij}\rho,E_{ij}^\ast\right]\right),\end{equation} where $E_{ij}$ are the standard basis elements of $M_N(\C)$ which have entry 1 in the $i$th row and $j$th column and all other entries are zero. We choose this terminology as given an digraph $G$ on $N$ vertices with weights $\gamma_{ij}$ one can consider the induced generator acting on $M_N(\C)$ given by \eqref{generic} for some appropriately chosen Hamiltonian $H$. Indeed, Rodr\'iguez-Rosario, Whitfield, and Aspuru-Guzik in \cite{RR} introduced such an example in the graph case (i.e. $\gamma_{ij}=\gamma_{ji}$) with $H=0$ to recover the classical random walk on $G$. Liu and Balu in \cite{LB}, also in the graph case, set $H$ to be the corresponding graph Laplacian (defined in Section~\ref{graphsection}) to give an alternate definition for a continuous-time open quantum random walk on $G$ (the original owing to Pellegrini in \cite{Pel}, and yet another by Sinayskiy and Petruccione in \cite{SP}); further, they show connected graphs induce uniquely relaxing semigroups. Glos, Miszczak, and Ostaszewski in \cite{Glos} extend this definition to digraphs by allowing $\gamma_{ij}\neq \gamma_{ji}$, and show $\mathcal{L}$ generates a uniquely relaxing semigroup for arbitrary $H$ if the digraph has strictly one terminal strongly connected component (defined in Section~\ref{digraphsection}).

In the case $H=\sum_{n=1}^Nh_nE_{nn}$ in \eqref{generic} we recover the generic generators, which were introduced (in the infinite dimensional case) by Accardi and Kozyrev in \cite{AK} as the stochastic limit of a discrete system with generic free Hamiltonian interacting with a mean zero, gauge invariant, 0-temperature, Gaussian field (and later generalized to positive temperature in \cite{AFH}). The finite-dimensional class of generic generators contain many well known and physically important models, such as coherent quantum control of a three-level atom in $\Lambda$-configuration interacting with two laser fields \cite{AKP}. Though the physical models require relations between the coefficients beyond what we write here, e.g. that $H$ is generic (hence the name), we ignore such restrictions and consider more generally any generator of form \eqref{generic} with $H=\sum_{n=1}^Nh_nE_{nn}$ a generic generator.

The generic generators are well studied and, though typically parsed in the language of Markov chains, some relations to digraph theory are known. Notably, from Accardi, Fagnola, and Hachica in \cite{AFH} it is known that given any matrix its diagonal and off-diagonal evolve independently of each other under the QDS arising from a generic generator, and in fact the action on diagonal operators describes the evolution of a classical continuous time Markov chain (with rates $\gamma_{ij}$) and the action on off-diagonal operators is given by conjugation with a contraction semigroup and its adjoint. With this relationship to Markov chains, Carbone, Sasso, and Umanita in \cite{CSU} find the general structure of the states fixed by the QDS, which can be computed given the kernel of the generator of the associated Markov chain. In that paper, these authors also examine the related problem of fixed points for the dual semigroup (Heisenberg picture) in context of the decoherence-free subalgebra (see also \cite{Evans, Dhahri,Carbone,Carbone2} and references therein).

The purpose of this work is twofold: First, we generalize the digraph induced generators given by \eqref{generic} in such a way that the results mentioned above remain true. We accomplish this generalization by allowing additional interaction coefficients, such as $\gamma_{ii}$, which preserve the main structural properties (notably, that if the Hamiltonian is diagonal then the diagonal and off-diagonal of a matrix evolve independently). We call such generators `pair block diagonal' generators, for reasons which will be made clear, and compute explicitly all invariant states in the diagonal Hamiltonian case. Second, we consider the converse construction to show that every QDS generator naturally gives rise to a digraph, and that under certain assumptions the properties of this digraph can be exploited to gain knowledge of both the number and the structure of the invariant states of the corresponding semigroup.

\subsection{Structure}

The structure of this article is as follows:

\noindent$\bullet$ In Section~\ref{GKSL2section} we establish formal definitions and notation for QDSs, and then provide a characterization of when the GKSL form defines a proper generator when allowed arbitrary orthonormal Lindblad operators.A physical three-level system is discussed to highlight some differences between the forms. In Section~\ref{contractivitysection} we note the equivalence between identity preservation and contractivity of a QDS in some, equivalently all Schatten $p$-norms for $p>1$.

\noindent$\bullet$ In Section~\ref{E basis} we establish the bulk of our notation and examine the structural properties of a generator when written with respect to the standard basis, which allows us to motivate and define the class of pair block diagonal generators (which contains the aforementioned digraph induced generators). Whereas the digraph induced generators can be used to model jumps between vector states, we remark that the pair block diagonal generators can be used to model jumps between superpositions of states. In Section~\ref{GM basis} we rephrase this notation and definition in terms of the Gell-Mann basis.

\noindent$\bullet$ In Sections~\ref{graphsection} and \ref{digraphsection} we establish the necessary graph and digraph terminology, as well as recall the necessary results.

\noindent$\bullet$ In Section~\ref{maindigraphsection} we define our main digraph of interest and show explicitly that every generator is naturally associated to a digraph through restriction to the diagonal subalgebra of $M_N(\C)$. We explicitly give the kernel of such restrictions.

\noindent$\bullet$ In Section~\ref{section LO} we consider the action of pair block diagonal generators on the off-diagonal subspace, and compute explicitly the eigenvalues and eigenmatrices of such. In Section~\ref{section L} we combine these kernel representations of the diagonal and off-diagonal restrictions to give an explicit formula for the kernel of a pair block diagonal generator, and thereby an explicit formula for all invariant states of the corresponding QDS.

\noindent$\bullet$ In Section~\ref{maxdisssection} we examine QDSs which are contractive for Schatten norms $p>1$ and show all invariant states of such QDSs are invariant for a naturally associated graph induced QDS. In Section~\ref{consistentsection} we define the notion of consistent generators as those which have Hamiltonian consistent with the naturally associated digraph, and show such generators have a lower bound on the number of invariant states for the corresponding QDS based on the connectedness of the digraph.

\section{General Properties of QDSs}

\subsection{The Form of $\mathcal{L}$}\label{GKSL2section}

Formally, a QDS (in the Schr\"odinger picture) on $M_N(\C)$ is a one-parameter family of operators $(T_t)_{t\geq 0}$ of $M_N(\C)$ satisfying: \begin{itemize}
	\item $T_0$ is the identity on $M_N(\C)$,
	\item $T_{t+s}=T_tT_s$ for all $t,s\geq0$,
	\item $t\mapsto T_t(A)$ is (weakly) continuous for all $A\in M_N(\C)$,
	\item $\Tr(T_t(A))=\Tr(A)$ for all $A\in M_N(\C)$ and all $t\geq 0$, and
	\item $T_t$ is completely positive for all $t\geq0$.
\end{itemize} Let $D_N(\C)$ denote the set of $N\times N$ states (i.e. positive semidefinite matrices of unit trace). When restricted to $D_N(\C)$ the QDS describes the Schr\"odinger dynamics of a quantum system with finitely many degrees of freedom. Every QDS on $M_N(\C)$ can be written in the form $T_t=e^{t\mathcal{L}}:=\sum_{k=0}^\infty t^k\mathcal{L}^k/k!$, where $\mathcal{L}(x)=\lim_{t\downarrow 0}\frac{1}{t}(T_t(x)-x)$ is called the generator of the QDS. Let $S_2^N$ denote $M_N(\C)$ endowed with the norm $||A||_2=(\Tr(|A|^2))^{1/2}$, which is induced by the Hilbert-Schmidt inner product $\langle A,B\rangle =\Tr(A^*B)$. The following characterization of such $\mathcal{L}$ is the renowned GKSL form:

\begin{theorem}[\cite{GKS,L}] \label{GKSL} Let $\{F_i|1\leq i \leq N^2-1\}$ be a set of $N\times N$ traceless orthonormal matrices (w.r.t. the Hilbert-Schmidt inner product). An operator $\mathcal{L}:M_N(\C)\to M_N(\C)$ is the generator of a QDS on $M_N(\C)$ if and only if it can be expressed in the form \begin{equation} \label{GKSLeq} \mathcal{L}(\rho)=-\imath[H,\rho]+\frac{1}{2}\sum_{i,j=1}^{N^2-1}c_{ij}([F_i,\rho F_j^\ast]+[F_i\rho,F_j^\ast]),\end{equation} with $H$ Hermitian and $C=(c_{ij})$ an $(N^2-1)\times (N^2-1)$ positive semidefinite matrix. Given $\mathcal{L}$ the Hamiltonian $H$ is uniquely determined by $\Tr(H)=0$; given $\mathcal{L}$ the coefficient matrix $C$ is uniquely determined by the choice of $F_i$'s.
\end{theorem}

If $H=0$ we say $\mathcal{L}$ is Hamiltonian-free. We note that $H$ describes the reversible dynamics of the system, and that all physically important information pertaining to the irreversible dynamics is contained in the positive semidefinite matrix $C$.

We are particularly interested in characterizing invariant states of a given QDS $(T_t)_{\geq 0}$; that is, states $\rho\in D_N(\C)$ satisfying $T_t(\rho)=\rho$ for all $t\geq 0$. To this end, notice that if $T_t(x)=x$ for all $t\geq0$ then $\mathcal{L}(x)=\lim_{t\downarrow0}\frac{1}{t}(T_t(x)-x)=0$, and if $\mathcal{L}(x)=0$ then certainly $T_t(x)=\sum_{k=0}^\infty t^k\mathcal{L}^k(x)/k!=x$. Hence a $T_t(x)=x$ for all $t\geq 0$ if and only if $\mathcal{L}(x)=0$. Recalling Lemma~17 of \cite{BN}, which states that $\ker\mathcal{L}$ is spanned by states, we have \begin{equation} \label{kerstates}\ker \mathcal{L}=\Span\{\rho\in D_N(\C):T_t(\rho)=\rho\text{ for all } t\geq0\}.\end{equation}

Note that $\dim\ker\mathcal{L}\geq 1$ since $\mathcal{L}$ has traceless range, and so every QDS possesses at least one invariant state.

Let $M_N^0(\C)$ denote the set of $N\times N$ traceless matrices. Given two orthonormal bases $\{F_i|1\leq i \leq N^2-1\}$ and $\{G_i|1\leq i \leq N^2-1\}$ of $M_N^0(\C)$ there is an $(N^2-1)\times(N^2-1)$ unitary matrix $U$ such that $[G_1, G_2,\ldots, G_{N^2-1}]=[F_1, F_2,\ldots, F_{N^2-1}]U$, representing the change of basis from $G_i$'s to $F_i$'s; that is, for $U=(u_{ij})$, we have $G_i=\sum_{k=1}^{N^2-1}u_{ki}F_k$ and contrariwise $F_{i}=\sum_{k=1}^{N^2-1}\ol{u}_{ik}G_k$ for all $1\leq i\leq N^2-1$. Considering \eqref{GKSLeq}, we have $\mathcal{L}(\rho)+i[H,\rho]=$ \begin{align*}
	&= \frac{1}{2}\sum_{i,j=1}^{N^2-1}c_{ij}([F_i,\rho F_j^\ast]+[F_i\rho,F_j^\ast]) \\ &= \frac{1}{2}\sum_{i,j=1}^{N^2-1}c_{ij}\left(\left[\sum_{k=1}^{N^2-1}\ol{u}_{ik}G_k,\rho \left(\sum_{\ell=1}^{N^2-1}\ol{u}_{ j\ell }G_\ell\right)^\ast\right]+\left[\sum_{k=1}^{N^2-1}\ol{u}_{ik}G_k\rho,\left(\sum_{\ell=1}^{N^2-1}\ol{u}_{j\ell}G_\ell\right)^\ast\right]\right) \\ &= \frac{1}{2}\sum_{i,j,k,\ell=1}^{N^2-1}\ol{u}_{ik}c_{ij}u_{j\ell }\left(\left[G_k,\rho G_\ell^\ast\right]+\left[G_k\rho,G_\ell^\ast\right]\right) \\ &=\frac{1}{2}\sum_{k,\ell=1}^{N^2-1}\widetilde{c}_{k\ell}\left(\left[G_k,\rho G_\ell^\ast\right]+\left[G_k\rho,G_\ell^\ast\right]\right) ,
	\end{align*} where $\widetilde{c}_{k\ell}=\sum_{i,j=1}^{N^2-1}\ol{u}_{ik}c_{ij}u_{j\ell}$ are the entries of $\widetilde{C}=U^\ast CU$. Thus, the $(N^2-1)\times (N^2-1)$ matrix $C$ when viewed as an operator $C:M_N^0(\C)\to M_N^0(\C)$ is uniquely determined by $\mathcal{L}$, with the choice of $F_i$'s being nothing but a choice of which orthonormal basis of $M_N^0(\C)$ for the matrix form of $C$ to be represented in.

	This operator viewpoint allows us to view every QDS generator $\mathcal{L}$ as the pair $H$ and $C$ uniquely determined by Theorem~\ref{GKSL}. If we drop the traceless requirement from Theorem~\ref{GKSL} so that the coefficient matrix acts on all of $M_N(\C)$ instead of just $M_N^0(\C)$, then we need to require stronger operator level properties (i.e., properties that do not rely on the choice of basis) to guarantee $\mathcal{L}$ is a QDS generator.
	
	\begin{theorem}\label{GKSL2} Let $\{F_i|1\leq i \leq N^2\}$ be a set of $N\times N$ orthonormal matrices (w.r.t. the Hilbert-Schmidt inner product). An operator $\mathcal{L}:M_N(\C)\to M_N(\C)$ is the generator of a QDS on $M_N(\C)$ if and only if it can be expressed in the form \begin{equation} \label{GKSL2eq} \mathcal{L}(\rho)=-\imath[\widetilde H,\rho]+\frac{1}{2}\sum_{i,j=1}^{N^2}\gamma_{ij}([F_i,\rho F_j^\ast]+[F_i\rho,F_j^\ast]),\end{equation} with $\widetilde H$ Hermitian and $\Gamma=(\gamma_{ij})$ an $N^2\times N^2$ matrix, regarded as acting on $M_N(\C)$ equipped with basis $\{F_i\}$, satisfying \begin{itemize}
			\item $P\Gamma |_{M_N^0(\C)}\geq 0$, where $P$ is the orthogonal projection from $M_N(\C)$ onto $M_N^0(\C)$, and
			\item $\re\Tr(\Gamma (A))=\re\Tr(\Gamma (I_N)A)$ for all Hermitian $A\in M_N(\C)$.
		\end{itemize}
	The operator $P\Gamma |_{M_N^0(\C)}$ is uniquely determined by $\mathcal{L}$. These conditions are satisfied if $\Gamma\geq 0$. 
	\end{theorem}

We remark that Theorem~\ref{GKSL2} is a natural extension of Theorem~\ref{GKSL}, in that the latter can be recovered by defining operator $\Gamma:M_N(\C)\to M_N(\C)$ by $\Gamma|_{M_N^0(\C)}=C$ and $\Gamma(I_N)=0$. Indeed, in this case $P\Gamma |_{M_N^0(\C)}=C\geq0$ and $\Tr(\Gamma(A))=0$ for all $A\in M_N(\C)$ simply because $C$ has traceless range.

\begin{proof} As \eqref{GKSLeq} is a special case of \eqref{GKSL2eq}, it suffices to prove that \eqref{GKSL2eq} always defines as QDS generator. Since the preceding argument for converting bases did not rely on any properties of the $F_i$'s or $G_i$'s beyond orthonormality, it will suffice to prove this for a fixed orthonormal basis $\{F_i\}$. To this end, we assume without loss of generality that $F_{N^2}=I_N/\sqrt{N}$ and that each $F_i$ is Hermitian (e.g., the Gell-Mann basis defined in Section~\ref{GM basis}). First note that the value of $\gamma_{N^2N^2}$ has no effect on the action of $\mathcal{L}$, since $\gamma_{N^2N^2}([I_N/\sqrt{N},\rho I_N/\sqrt{N}]+[I_N/\sqrt{N} \rho,I_N/\sqrt{N}])=0$. We thus assume that $\gamma_{N^2N^2}=0$. Next, we compute \begin{align*} &\phantom{=}\gamma_{iN^2}\left(\left[F_i,\rho\frac{I_N}{\sqrt{N}}\right]+\left[F_i \rho,\frac{I_N}{\sqrt{N}}\right]\right)+\gamma_{N^2i}\left(\left[\frac{I_N}{\sqrt{N}},\rho F_i\right]+\left[\frac{I_N}{\sqrt{N}} \rho F_i\right]\right)=\\
		&=\frac{\gamma_{iN^2}}{\sqrt{N}}[F_i,\rho]+\frac{\gamma_{N^2i}}{\sqrt{N}}[\rho, F_i] = \frac{\gamma_{iN^2}-\gamma_{N^2i}}{\sqrt{N}}[F_i,\rho] = -{\imath}\left[\frac{\im (\gamma_{N^2i}-\gamma_{iN^2})}{\sqrt{N}}F_i,\rho\right]
		\end{align*} where the last equality follows since $$\re\gamma_{iN^2} =\re\Tr\left(F_i\Gamma\left(\frac{I_N}{\sqrt{N}}\right)\right)=\re\Tr\left(\Gamma\left(F_i\right)\frac{I_N}{\sqrt{N}}\right)=\re\gamma_{N^2i}$$ by assumption. Thus the real parts of these coefficients have no effect on the action of $\mathcal{L}$, so we may assume $\re\gamma_{iN^2}=\re\gamma_{N^2i}=0$ for all $i=1,\ldots,N^2-1$. Further, since the imaginary parts act as a commutator, we may write \begin{equation}\label{H shift}\mathcal{L}=-\imath\left[\widetilde{H}+\sum_{i=1}^{N^2-1}\frac{\im (\gamma_{N^2i}-\gamma_{iN^2})}{2\sqrt{N}}F_i,\rho\right]+\frac{1}{2}\sum_{i,j=1}^{{N^2-1}}\gamma_{ij}([F_i,\rho F_j^\ast]+[F_i\rho,F_j^\ast]),\end{equation} which is of GKSL form \eqref{GKSLeq} since $P\Gamma |_{M_N^0(\C)}=(\gamma_{ij})_{i,j=1}^{N^2-1}\geq 0$ and each $F_i$ Hermitian implies $H=\widetilde{H}+\sum_{i=1}^{N^2-1}\frac{\im (\gamma_{N^2i}-\gamma_{iN^2})}{2\sqrt{N}}F_i$ is Hermitian. Uniqueness of the operator $P\Gamma |_{M_N^0(\C)}$ also follows from Theorem~\ref{GKSL}.
	
It remains to show that these conditions are satisfied if $\Gamma\geq0$. That $P\Gamma|_{M_N^0(\C)}\geq0$ follows immediately since every principal submatrix of a positive semidefinite matrix is positive semidefinite (consider the quadratic form $\Tr(A^\ast \Gamma(A))\geq 0$ restricted to traceless $A$). That $\re\Tr(\Gamma (A))=\re\Tr(\Gamma (I_N)A)$ for $A$ Hermitian (in $S_2^N$) follows since $\Gamma$ is Hermitian (on $S_2^N$). Explicitly, \[\Tr(\Gamma (I_N)A)=\Tr(A\Gamma (I_N))=\langle A,\Gamma(I_N)\rangle=\langle \Gamma(A),I_N\rangle=\ol{\langle I_N,\Gamma(A)\rangle}=\ol{\Tr(\Gamma(A))},\] and so $\re\Tr(\Gamma(I_N)A)=\re\overline{\Tr(\Gamma(A))}=\re\Tr(\Gamma(A))$. \end{proof}

We ward here against the thought that allowing the matrices $F_i$ to have trace in GKSL form \eqref{GKSLeq} equates to `shifting' some of the action of $-\imath[H,\cdot]$ to the dissipative part (i.e., $\mathcal{L}+\imath[H,\cdot]$). That indeed is the case in the previous proof, but this relied on our choice of $F_i$'s being both traceless and Hermitian. For general $F_i$'s the interaction is more subtle, and indeed it is easy to construct examples of Hamiltonian-free $\mathcal{L}$ written in GKSL form \eqref{GKSLeq} which are equivalent to Hamiltonian-free form \eqref{GKSL2eq} with only $F_i$'s of unit trace appearing ($\mathcal{L}_d$ defined in Example~\ref{rydberg} at the end of this subsection is one such example).

What is true, however, is that one can disallow any `shifting' of the action of $-\imath[H,\cdot]$ to the dissipative part by choosing $\widetilde{H}$ to be $H$ uniquely determined by Theorem~\ref{GKSL}, and $\Gamma$ to be the natural dilation of the operator $C$ uniquely determined by Theorem~\ref{GKSL}.

\begin{theorem}\label{GKSL3} Let $\{F_i|1\leq i \leq N^2\}$ be a set of $N\times N$ orthonormal matrices (w.r.t. the Hilbert-Schmidt inner product). An operator $\mathcal{L}:M_N(\C)\to M_N(\C)$ is the generator of a QDS on $M_N(\C)$ if and only if it can be expressed in the form \begin{equation} \label{GKSL3eq} \mathcal{L}(\rho)=-\imath[H,\rho]+\frac{1}{2}\sum_{i,j=1}^{N^2}\gamma_{ij}([F_i,\rho F_j^\ast]+[F_i\rho,F_j^\ast]),\end{equation} with $H$ traceless and Hermitian, and $\Gamma=(\gamma_{ij})$ an $N^2\times N^2$ matrix, regarded as acting on the basis $\{F_i\}$, satisfying \begin{itemize}
		\item $\Gamma \geq 0$,
		\item $\Gamma(I_N)=0$, and
		\item $\Tr(\Gamma(A))=0$ for all $A\in M_N(\C)$.
	\end{itemize}
	Given $\mathcal{L}$ the Hamiltonian $H$ is uniquely determined by $\Tr(H)=0$ (and is the same as $H$ as Theorem~\ref{GKSL}); given $\mathcal{L}$ the coefficient matrix $\Gamma$ is uniquely determined by the choice of $F_i$'s.
\end{theorem}

\begin{proof}
As before, given QDS generator $\mathcal{L}$ we may write it it form \eqref{GKSLeq} with any traceless orthonormal basis $\{\widetilde{F}_i\}$ and define $\Gamma:M_N(\C)\to M_N(\C)$ by $\Gamma|_{M_N^0(\C)}=C$ and $\Gamma(I_N)=0$. Changing the basis from $\{\widetilde{F}_i\}$ to the desired $\{F_i\}$ preserves the operator properties $\Gamma \geq 0$, $\Gamma(I_N)=0$, and $\Tr(\Gamma(A))=0$, and the coefficients of the resulting matrix are uniquely determined by this basis change. The converse is a special case of Theorem~\ref{GKSL2}.
\end{proof}

Though easier to check as compared to Theorem~\ref{GKSL2}, the disadvantage of Theorem~\ref{GKSL3} is that one may fail to detect if a given equation represents a QDS generator in the case $\Gamma$ fails to satisfy these stronger properties. The following example illustrates this, as well as the importance of allowing the $F_i$'s to have trace when considering phenomenological operators.

\begin{example}\label{rydberg} We follow \cite{Garttner,Pritchard, Hofmann}, and consider a single three-level atom with ground, excited, and Rydberg states $$|g\rangle =\left(\begin{smallmatrix}
	1 \\ 0 \\ 0 
	\end{smallmatrix}\right),\qquad|e\rangle =\left(\begin{smallmatrix}
	0 \\ 1 \\ 0 
	\end{smallmatrix}\right),\qquad |r\rangle =\left(\begin{smallmatrix}
	0 \\ 0 \\ 1 
	\end{smallmatrix}\right),$$ interacting with two laser fields: a probe laser field which drives the transition from the ground to the excited state, and a coupling laser field which drives the transition from the excited to the Rydberg state. In this regime there are two decay modes: one from $|e\rangle$ to $|g\rangle$ at rate $\Gamma_{eg}$, and another from $|r\rangle$ to $|e\rangle$ at rate $\Gamma_{re}$. The spontaneous emission from $|a\rangle$ to $|b\rangle$ is described by setting $F_i=F_j=\sqrt{\Gamma_{ab}}|b\rangle\langle a|$ in \eqref{GKSLeq}; that is, by the GKSL operator $$\mathcal{L}_{ab}(\rho)=\Gamma_{ab}([\,|b\rangle\langle a|\rho,|a\rangle\langle b|\,]+[\,|b\rangle\langle a|,\rho|a\rangle\langle b|\,]).$$ Due to the finite linewidths of the laser fields, there are additional dephasing mechanisms which lead to additional decay of the coherences between states. The line width of the laser driving a transition from $|a\rangle$ to $|b\rangle$ can be taken into account by phenomenological operator $$\mathcal{L}_{ab}^d(\rho)=-\frac{\Gamma_{ab}^d}{2}(|a\rangle\langle a| \rho |b\rangle\langle b|+|b\rangle\langle b|\rho|a\rangle\langle a|),$$ where $\Gamma_{ab}^d$ is the full width of the spectral laser profile. Note that such operators are not of GKSL type, but they can be written as a linear combination of GKSL operators via $$\mathcal{L}_{ab}^d=\frac{\Gamma_{ab}^d}{2}(\mathcal{L}_{aa}+\mathcal{L}_{bb}-\mathcal{L}_{cc}),$$ where $(a,b,c)$ are permutations of $(g,e,r)$ and $\Gamma_{aa}=\Gamma_{bb}=\Gamma_{cc}=1$. In total, the master equation describing the system is given by $$\partial_t\rho=\mathcal{L}(\rho)=-\imath[H,\rho]+\mathcal{L}_{eg}(\rho)+\mathcal{L}_{re}(\rho)+\mathcal{L}_{ge}^d(\rho)+\mathcal{L}_{er}^d(\rho)+\mathcal{L}_{gr}^d(\rho),$$ where $H$ describes the time evolution in the absence of decoherence. We focus on the extra dephasing terms, and define $$\mathcal{L}_d=\mathcal{L}_{ge}^d+\mathcal{L}_{er}^d+\mathcal{L}_{gr}^d=\frac{1}{2}\left((\Gamma_{ge}^d+\Gamma_{gr}^d-\Gamma_{er}^d)\mathcal{L}_{gg}+(\Gamma_{ge}^d+\Gamma_{er}^d-\Gamma_{gr}^d)\mathcal{L}_{ee}+(\Gamma_{gr}^d+\Gamma_{er}^d-\Gamma_{ge}^d)\mathcal{L}_{rr}\right).$$ Consider the diagonal subalgebra $\mathcal{D}=\Span(|g\rangle\langle g|, |e\rangle \langle e|,|r\rangle\langle r|)$ of $M_N(\C)$. Since $\mathcal{L}_d|_\mathcal{D}=0$ it is tempting to write that $\mathcal{L}_d$ cannot be written in GKSL form \eqref{GKSLeq} (see e.g. section 4.1.1 of \cite{Pritchard}). Regarding the coefficient matrix $\Gamma$ of $\mathcal{L}_d$ as acting of $M_N(\C)$, however, we have that $\Gamma|_{\mathcal{D}^\perp}=0$ and $\Gamma|_{\mathcal{D}}:\mathcal{D}\to\mathcal{D}$ acts by $$\Gamma|_{\mathcal{D}}=\frac{1}{2} \begin{pmatrix}
	\Gamma_{ge}^d+\Gamma_{gr}^d-\Gamma_{er}^d &  &  \\ 
	& \Gamma_{ge}^d+\Gamma_{er}^d-\Gamma_{gr}^d &  \\ 
	&  & \Gamma_{gr}^d+\Gamma_{er}^d-\Gamma_{ge}^d
	\end{pmatrix}.$$ This matrix is Hermitian and under mild conditions positive semidefinite (e.g. consider independent lasers, so that $\Gamma_{gr}=\Gamma_{ge}+\Gamma_{er}$). In such a case it is immediate that $\Gamma$ satisfies the conditions of Theorem~\ref{GKSL2}, and so $\mathcal{L}_d$ is indeed a GKSL generator. Because the summation of operators of form \eqref{GKSLeq} returns another operator of that form, this implies $\mathcal{L}$ itself is a GKSL operator.

Note that $\mathcal{L}_d$ is a Hamiltonian-free QDS generator in form \eqref{GKSL2eq} with only $F_i$'s of unit trace appearing. The given representation is not of form \eqref{GKSL3eq}, however, as $\Gamma$ has not been chosen properly to satisfy the stronger conditions of Theorem~\ref{GKSL3}. To write $\mathcal{L}$ in form \eqref{GKSL3eq} we replace $\Gamma|_{\mathcal{D}}$ above by $$\widetilde{\Gamma|_{\mathcal{D}}}=\frac{1}{18} \begin{pmatrix}
	4\Gamma_{ge}+4\Gamma_{gr}-2\Gamma_{er} & \Gamma_{gr}-5\Gamma_{ge}+\Gamma_{er} & \Gamma_{ge}-5\Gamma_{gr}+\Gamma_{er} \\ 
	\Gamma_{gr}-5\Gamma_{ge}+\Gamma_{er} & 4\Gamma_{ge}-2\Gamma_{gr}+4\Gamma_{er} & \Gamma_{ge}+\Gamma_{gr}-5\Gamma_{er} \\ 
	\Gamma_{ge}-5\Gamma_{gr}+\Gamma_{er} & \Gamma_{ge}+\Gamma_{gr}-5\Gamma_{er} & 4\Gamma_{gr}-2\Gamma_{ge}+4\Gamma_{er}
	\end{pmatrix},$$ which can be found by writing $\Gamma$ in terms of a Hermitian orthonormal basis $\{F_i|1\leq i \leq 9\}$ with $F_1,\ldots, F_8$ traceless and $F_9=I_3/\sqrt{3}$ as in the proof of Theorem~\ref{GKSL2}, setting equal to zero the non-contributing terms (i.e., setting $\gamma_{99}=\re\Gamma_{i9}=\re\gamma_{9i}=0$ for all $i=1,\ldots,8$), and then rewriting $\Gamma$ again back in terms of the original basis. Because $H=0$, and forms \eqref{GKSLeq} and \eqref{GKSL3eq} use the same Hamiltonian, any representation of $\mathcal{L}_d$ in form \eqref{GKSLeq} is Hamiltonian-free. In particular, allowing the matrices $F_i$ to have trace in GKSL form \eqref{GKSLeq} is not equivalent to `shifting' some of the action of $-\imath[H,\cdot]$ to the dissipative part (i.e., $\mathcal{L}+\imath[H,\cdot]$).
\end{example}

\subsection{Contractivity of $\boldsymbol{T_t}$}\label{contractivitysection}

For $1\leq p \leq \infty$, we call $M_N(\C)$ endowed with the Schatten $p$-norm $||A||_p=(\Tr(|A|^p))^{1/p}$ for $p<\infty$ and $||A||_\infty=\sup_{||v||=1}||Av||$ the $p$-Schatten space $S_p^N$. In particular, $S_2^N$ is the Hilbert-Schmidt space defined previously and $S_1^N$ is the usual trace class space. For $T:M_N(\C)\to M_N(\C)$, let $||T||_{p\to p}$ denote the operator norm $||T||_{p\to p}=\sup _{x\in M_N(\C)}\frac{||T(A)||_{p}}{||A||_{p}}$.

It is well known that every QDS $(T_t)_{t\geq 0}$ is a contraction semigroup on $S_1^N$ (i.e., satisfies $||T_t||_{1\to 1}\leq1$ for all $t\geq 0$). Indeed, if $T$ is trace preserving and positive then its trace-dual $T^\dagger$ is unital and positive, and hence achieves its norm at the identity. Thus, $||T||_{1\to 1}=||T^\dagger||_{\infty\to\infty}=||T^\dagger(I_N)||_{\infty}=||I_N||_\infty=1$ (actually, if $T$ is trace preserving then $||T||_{1\to 1}\leq1$ if and only if $T$ is positive; see Proposition~2.11 of \cite{P}). We wish to take advantage of the Hilbert space properties of $S_2^N$, however, so we seek QDSs which are contraction on $S_2^N$. The Lumer-Phillips Theorem states that $||T_t||_{2\to 2}\leq 1$ for all $t$ if and only if the generator $\mathcal{L}$ satisfies $\re \Tr(x^*\mathcal{L}(x))\leq 0$ for all $x\in M_N(\C)$ (see e.g. Corollary II.3.20 of \cite{EN}). We particularize a result of P\'erez-Garc\'ia, Wolf, Petz, and Ruskai \cite{PG} to offer the following characterization, and compare it to this well known Lumer-Phillips result:

\begin{corollary} \label{diss} Suppose $(T_t)_{t\geq 0}$ is a QDS with generator $\mathcal{L}$. The following are equivalent: \begin{itemize}
		\item $||T_t||_{p\to p}\leq 1$ for some $1<p\leq \infty$ and all $t\geq 0$,
		\item $||T_t||_{p\to p}\leq 1$ for all $1\leq p\leq \infty$ and all $t\geq 0$,
		\item $\mathcal{L}(I_N)=0$. 
	\end{itemize}
	In this case $\Tr(x\mathcal{L}(x))\leq 0$ for all Hermitian matrices $x\in M_N(\C)$.
\end{corollary}

\begin{proof} Considering fixed $t$, we have that $||T_t||_{p\to p}\leq 1$ for some, equivalently all $1<p\leq \infty$ if and only if $T_{t}(I_N)=I_N$ by Theorem~II.4 of \cite{PG}. The result then follow from \eqref{kerstates}, which shows $T_t(I_N)=I_N$ for all $t\geq 0$ if and only if $\mathcal{L}(I_N)=0$, as desired.
	
	For the second statement, since the Lumer-Phillips Theorem gives that $\re \Tr(x^*\mathcal{L}(x))\leq 0$ for all $x\in M_N(\C)$, it suffices to prove that $\Tr(x\mathcal{L}(x))\in \R$ for Hermitian $x$. This follows immediately from \[\overline{\Tr(x\mathcal{L}x)}=\Tr((x\mathcal{L}(x))^\ast)=\Tr(\mathcal{L}(x)^\ast x^\ast)=\Tr(x^\ast\mathcal{L}(x)^\ast)=\Tr(x^\ast\mathcal{L}(x^\ast))=\Tr(x\mathcal{L}(x)),\] where we use that $\mathcal{L}(x)^\ast=\mathcal{L}(x^\ast)$ since $T(x)^\ast=T(x^\ast)$ (as a positive linear map). \end{proof}

One may read the previous Corollary as saying a QDS is contractive for all Schatten $p$-norms if and only if the maximally mixed state $I_N/N$ is invariant. Calling an operator $T:M_N(\C)\to M_N(\C)$ Hermitian if it is Hermitian when regarded as $T:S_2^N\to S_2^N$, the next result describes potential invariant states of such a QDS given a Hermitian `part' of its generator.

\begin{lemma} \label{L=A+B}
	Suppose $\mathcal{L}$ is a QDS generator satisfying $\mathcal{L}(I_N)=0$ which can be written $\mathcal{L}=\mathcal{A}+\mathcal{B}$ with $\mathcal{A}$ and $\mathcal{B}$ each a QDS generator. If $\mathcal{A}$ is Hermitian and $\mathcal{A}(I_N)=0$ then $\ker \mathcal{L}\subseteq \ker \mathcal{A}$.
\end{lemma}

\begin{proof} Since \eqref{kerstates} shows that $\ker\mathcal{L}$ is spanned by states, it suffices to show that if $\mathcal{L}(\rho)=0$ for some state $\rho$ then $\mathcal{A}(\rho)=0$. To this end, notice that $\mathcal{A}(I_N)=\mathcal{L}(I_N)=0$ implies $\mathcal{B}(I_N)=0$, and so $\Tr(x \mathcal{A}(x))\leq 0$ and $\Tr(x \mathcal{B}(x))\leq 0$ for all Hermitian $x$ by Corollary~\ref{diss}. Fixing state $\rho$ such that $\mathcal{L}(\rho)=0$, equivalently $\mathcal{A}(\rho)=-\mathcal{B}(\rho)$, we must then have $\Tr(\rho \mathcal{A}(\rho))=0$. Thus, \[-\Tr(\rho\mathcal{A}(\rho))=\langle \rho, -\mathcal{A}\rho \rangle = \langle (-\mathcal{A})^{1/2}\rho,(-\mathcal{A})^{1/2}\rho\rangle =0\] implies $(-\mathcal{A})^{1/2}\rho =0$, and hence $\mathcal{A}\rho =0$.
\end{proof}

\section{The Matrix Representation of $\mathcal{L}$}

\subsection{The Standard Basis} \label{E basis}

Our proofs rely on exact calculations and the ability to move between two well-known bases of $M_N(\C)$: the standard basis and the (generalized) Gell-Mann basis (introduced in Section~\ref{GM basis}). Recall that the standard basis consists of the $N\times N$ matrices $E_{ij}$ that have entry 1 in the $i$th row and $j$th column and all other entries are zero. It is easy to see that the standard basis satisfies $E_{ij}E_{k\ell}=\delta_{jk}E_{i\ell}$, where $\delta_{jk}$ is the standard Kronecker delta.

By way of Theorem~\ref{GKSL2}, every QDS generator $\mathcal{L}$ can be written with respect to the standard basis; that is, \begin{equation} \label{E form} \mathcal{L}(\rho)=-\imath[\widetilde{H},\rho]+\frac{1}{2}\sum_{i,j,k,\ell=1}^N\gamma_{ijk\ell}\left(\left[E_{ij},\rho E_{k\ell}^\ast\right]+\left[E_{ij}\rho,E_{k\ell}^\ast\right]\right).\end{equation}

We henceforth reserve $\Gamma$ to denote the $N^2\times N^2$ coefficient matrix $\Gamma:=(\gamma_{ijk\ell})$ for $\mathcal{L}$ written with respect to the standard basis, and so always assume $\Gamma$ satisfies the criteria of Theorem~\ref{GKSL2}. We use \[D_{ijk\ell}:=[E_{ij},\cdot E_{ \ell k}]+[E_{ij}\cdot,E_{\ell k}]\] to denote the individual {\bf Lindblad operators} written with respect to the standard basis. For $(i,j)=(k,\ell)$, the so-called {\bf diagonal Lindblad operators}, we use the simplified notation \[D_{ij}:=[E_{ij},\cdot E_{ji}]+[E_{ij}\cdot,E_{ji}].\]

We are interested in matrix representations for $\Gamma$ and $\mathcal{L}$ with respect to the standard basis, and to this end we order the standard basis of $M_N(\C)$ by pairing together $E_{ij}$ and $E_{ji}$ for $i\neq j$, then adjoining the diagonal $E_{nn}$. For example, for $N=3$ we may take the natural ordering $E_{12},E_{21},E_{13},E_{31},E_{23},E_{32},E_{11},E_{22},E_{33}$, but the exact ordering of the $E_{ij},E_{ji}$ pairs or the $E_{nn}$ is immaterial.

With this ordering, consider $\Gamma:M_N(\C)\to M_N(\C)$ written as an $N^2\times N^2$ matrix. Denote by $\Gamma^{\mathcal{O}}$ the $N(N-1)$ order leading principal submatrix of $\Gamma$; that is, $\Gamma^{\mathcal{O}}:\mathcal{O}\to \mathcal{O}$ is the submatrix formed by the rows and columns corresponding to the {\bf off-diagonal subspace} $\mathcal{O}:=\Span\{E_{ij}\}_{i,j=1; i\neq j}^N$ of $M_N(\C)$. Further, denote by $\Gamma^{\mathcal{D}}:\mathcal{D}\to \mathcal{D}$ the complementary submatrix formed by the rows and columns corresponding to the {\bf diagonal subalgebra} $\mathcal{D}:=\Span\{E_{nn}\}_{n=1}^N$ of $M_N(\C)$. Then \begin{equation*}\Gamma=\begin{pmatrix}
\Gamma^{\mathcal{O}} & \ast \\ 
\ast & \Gamma^{\mathcal{D}}
\end{pmatrix}.\end{equation*}

Since $\Gamma$ satisfies $P\Gamma |_{M_N^0(\C)}\geq0$ we have $\Gamma^{\mathcal{O}}\geq0$, as every principal submatrix of a positive semidefinite matrix is itself positive semidefinite. For each fixed pair $i,j,$ with $i<j$, we call the $2\times 2$ sub-matrix of $\Gamma^{\mathcal{O}}$ consisting of the rows and columns corresponding to $E_{ij}$ and $E_{ji}$ the {\bf $\boldsymbol{ij}$ block}. Note that each $ij$ block is positive semidefinite. Similar to the language used when referring to the diagonal of a matrix or when a matrix is diagonal, we refer to the collection of all $ij$ blocks of $\Gamma^{\mathcal{O}}$ as the {\bf pair block diagonal} of $\Gamma^{\mathcal{O}}$, and if $\Gamma^{\mathcal{O}}$ has no nonzero entries outside of its pair block diagonal we say $\Gamma^{\mathcal{O}}$ is pair block diagonal. We denote the upper-right entry of the $ij$ block by $\gamma_{ijji}=:\alpha_{ij}+\imath \beta_{ij}$ (and thus the lower-left by $\gamma_{jiij}=:\alpha_{ij}-\imath \beta_{ij}$), where $\alpha_{ij},\beta_{ij}\in\R$. Denote the diagonal entries of $\Gamma$ by $\gamma_{ijij}=:\gamma_{ij}$, $\gamma_{jiji}=:\gamma_{ji}$, and $\gamma_{nnnn}=:\gamma_{nn}$ in the natural way, noting $\gamma_{ij},\gamma_{ji}\geq0$ since $\Gamma^{\mathcal{O}}\geq0$.

To illustrate these notations, the following is an example of a matrix $\Gamma$ in dimension $N=3$ for which $\Gamma^{\mathcal{O}}$ is pair block diagonal and $\Gamma^{\mathcal{D}}$ is diagonal: $$\Gamma=\left(\begin{smallmatrix}
\gamma_{12} & \alpha_{12}+\imath \beta_{12} &  &  &  &  &  & & \\ 
\alpha_{12}-\imath \beta_{12} & \gamma_{21} &  &  &  &  &  & & \\ 
&  & \gamma_{13} & \alpha_{13}+\imath \beta_{13} &  &  &  & & \\ 
&  & \alpha_{13}-\imath \beta_{13} & \gamma_{31} &  &  &  &  &\\ 
&  &  &  & \gamma_{23} & \alpha_{23}+\imath \beta_{23} &  & & \\ 
&  &  &  & \alpha_{23}-\imath \beta_{23} & \gamma_{32} &  & & \\ 
&  &  &  &  &  & \gamma_{11} &  &\\ 
&  &  &  &  &  &  & \gamma_{22} &\\
&  &  &  &  &  &  &  &  \gamma_{33}
\end{smallmatrix}\right)$$

\begin{remark}\label{superposition}
	Fix orthogonal vector states $|i\rangle$ and $|j\rangle$ and consider a system which transfers superposition state $|\psi \rangle = a|i\rangle +b|j\rangle$ to superposition state $|\phi\rangle=c|i\rangle+d|j\rangle$ with probability $\gamma$ over a very short evolution time $dt$. To construct a model for such a system we make use of a short time expansion of the Kraus operator sum representation $\rho'=\sum_{\alpha}K_\alpha(dt)\rho K_\alpha^\ast(dt)$ (see e.g. section IX of \cite{Lidar}). Setting $$F_{ij}:=\frac{c}{b}E_{ij}+\frac{d}{a}E_{ji}$$ so that $F_{ij}|\psi\rangle=|\phi\rangle$, we take Kraus operator $$K_1(dt)=\sqrt{\gamma dt}F_{ij}$$ to represent the transition. Normalization $\sum_\alpha K_\alpha^\ast(dt)K_\alpha(dt)=I_N$ up to order $O(dt)$ (to ensure the evolution is trace preserving) requires a second Kraus operator  $$K_2(dt)=I_N-\frac{1}{2}K_1^\ast(dt) K_1(dt).$$ Thus, we have that $$\rho'=K_1(dt)\rho K_1^\ast(dt)+K_2(dt)\rho K_2^\ast(dt)=\rho+\gamma dt([F_{ij},\rho F_{ij}^\ast]+[F_{ij}\rho,F_{ij}^\ast]).$$ Assuming the same Kraus representation works over all time, we arrive at the GKSL equation $$\mathcal{L}(\rho)= \lim_{dt\to 0}\frac{\rho'-\rho}{dt}=\gamma ([F_{ij},\rho F_{ij}^\ast]+[F_{ij}\rho,F_{ij}^\ast]).$$ Rewriting $\mathcal{L}$ in terms of the standard basis \eqref{E form}, the coefficient matrix $\Gamma$ has nonzero entries only in the $ij$ block, which is given by $$\Gamma_{ij}=\gamma\begin{pmatrix}
		\frac{c\ol{c}}{b\ol{b}} & \frac{c\ol{d}}{\ol{a} b} \\ \frac{\ol{c}d}{a\ol{b}} & \frac{d\ol{d}}{a\ol{a}}
		\end{pmatrix}.$$ Thus, while diagonal coefficient matrices can be interpreted as describing jumps between states $|i\rangle$ and $|j\rangle$ (as with the graph induced generators \eqref{generic}), the pair block diagonal coefficient matrices can describe jumps between two superpositions of states $|i\rangle$ and $|j\rangle$. A main result of this work is to characterize invariant states of QDS generators with such coefficient matrices (see Theorem~\ref{main pair block in E} and Example~\ref{superpositionexample}).
\end{remark}

Extending the submatrix notations to $\mathcal{L}:M_N(\C)\to M_N(\C)$ in the natural way, we write \begin{equation}\label{matrix L}\mathcal{L}=\begin{pmatrix}
\mathcal{L}^{\mathcal{O}} & \ast \\ 
\ast & \mathcal{L}^{\mathcal{D}} 
\end{pmatrix}.\end{equation} We note that Havel considered the entries of $\mathcal{L}$ when written as such an $N^2\times N^2$ matrix to recover the coefficients of $\Gamma$ in terms of Choi matrices (Proposition~12 of \cite{H}). We are interested in the other direction, however: how the coefficients of $\Gamma$ affect the action of $\mathcal{L}$.

Per the introduction, we seek generators $\mathcal{L}$ which gives rise to QDSs which {\bf evolve independently} on $\mathcal{D}$ and $\mathcal{O}$ in the sense that \begin{equation*} T_t(A)=T_t^\mathcal{O}(\textrm{diag}(A))+T_t^\mathcal{D}(A-\textrm{diag}(A)) \end{equation*} for all $A\in M_N(\C)$. Since exponentiation preserves block diagonal structure, if $\mathcal{D}$ and $\mathcal{O}$ are each invariant for $\mathcal{L}$ (equivalently $\ast=0$ in \eqref{matrix L}), then $e^{t\mathcal{L}}=T_t=\left(\begin{smallmatrix}
T_t^\mathcal{O} & 0  \\ 
0 & T_t^\mathcal{D}
\end{smallmatrix}\right)$, where $T_t^\mathcal{O}:=e^{t\mathcal{L}^{\mathcal{O}}}$ and $T_t^\mathcal{D}:=e^{t\mathcal{L}^{\mathcal{D}}}$. Conversely, if $(T_t)_{t\geq 0}$ evolves independently on $\mathcal{D}$ and $\mathcal{O}$, then necessarily $\mathcal{D}$ and $\mathcal{O}$ are each invariant for $T_t$ for all $t\geq 0$, and hence invariant for $\mathcal{L}$. We are thus seeking generators for which $\ast=0$ in \eqref{matrix L}.

As each entry of $\mathcal{L}$'s matrix representation is a linear combination of entries of $\widetilde{H}$ and $\Gamma$ as determined by \eqref{E form}, we can consider how each entry of $\Gamma$ contributes to various entries of $\mathcal{L}$. Explicitly, we compute 

\begin{align} 
\notag D_{ijk\ell}(E_{st}) &= [E_{ij},E_{st}E_{\ell k}]+[E_{ij}E_{st},E_{\ell k}] \\ &= \label{E calculation} 2E_{ij}E_{st}E_{\ell k}-E_{st}E_{\ell k}E_{ij}-E_{\ell k}E_{ij}E_{st} \\ \notag &= 2\delta_{js}\delta_{\ell t}E_{ik}-\delta_{\ell t}\delta_{ik}E_{sj}-\delta_{ik}\delta_{js}E_{\ell t}.
\end{align}

In particular,
\begin{equation*}D_{ij}(E_{k\ell}) = -(\delta_{jk}+\delta_{j\ell})E_{k\ell},\qquad D_{ijji}(E_{k\ell})=2\delta_{jk}\delta_{i\ell}E_{\ell k}\end{equation*}
and \begin{equation*}D_{iijj}(E_{k\ell})=(2\delta_{ik}\delta_{j\ell}-\delta_{ij}\delta_{ik}-\delta_{ij}\delta_{j\ell})E_{k\ell}.\end{equation*} Notably, entries of $\Gamma^{\mathcal{D}}$ and of the pair block diagonal of $\Gamma^{\mathcal{O}}$ contribute only to $\mathcal{L}^{\mathcal{D}}$ and to the pair block diagonal of $\mathcal{L}^{\mathcal{O}}$. If we assume the Hamiltonian is diagonal, that is $\widetilde{H}=\sum_{n=1}^Nh_nE_{nn}$, then we compute \[-\imath[\widetilde{H},E_{k\ell}] = -\imath\sum_{n=1}^Nh_n[E_{nn},E_{k\ell}]=-\imath(h_k-h_\ell)E_{k\ell},\] and see that entries of $\widetilde{H}$ contribute only to the diagonal of $\mathcal{L}^\mathcal{D}$. This gives us the following:

\begin{remark} \label{pair block remark} Let $\mathcal{L}$ be a QDS generator written with respect to the standard basis \eqref{E form} with Hamiltonian $\widetilde{H}=\sum_{n=1}^Nh_nE_{nn}$. If $\Gamma=\left(\begin{smallmatrix}\Gamma^{\mathcal{O}} & 0 \\ 0 & \Gamma^{\mathcal{D}}
	\end{smallmatrix}\right)$ with $\Gamma^{\mathcal{O}}$ pair block diagonal, then \begin{equation}\label{block L} \mathcal{L}=\begin{pmatrix}
	\mathcal{L}^{\mathcal{O}} & 0 \\ 
	0 & \mathcal{L}^{\mathcal{D}} 
	\end{pmatrix} \end{equation} with $\mathcal{L}^{\mathcal{O}}$ pair block diagonal; in this case, if $\Gamma^{\mathcal{O}}$ is diagonal then $\mathcal{L}^{\mathcal{O}}$ is diagonal.
\end{remark}

A partial converses are also true: no entry of $\widetilde{H}$ outside its diagonal and no entry of $\Gamma$ outside both $\Gamma^\mathcal{D}$ and the pair block diagonal of $\Gamma^\mathcal{O}$ contributes to the pair block diagonal of $\mathcal{L}^{\mathcal{O}}$ or to $\mathcal{L}^{\mathcal{D}}$.

\begin{definition} \label{convention}
	We call QDS generator $\mathcal{L}$ {\bf pair block diagonal with respect to the standard basis} if $\mathcal{L}$ is of form \eqref{E form} with \begin{equation*} \Gamma=\begin{pmatrix} \Gamma^{\mathcal{O}} & 0 \\ 0 & \Gamma^{\mathcal{D}}\end{pmatrix}\end{equation*} and $\Gamma^{\mathcal{O}}$ pair block diagonal.
\end{definition}

Note that a generator which is pair block diagonal with respect to the standard basis with $\widetilde{H}=\sum_{n=1}^Nh_nE_{nn}$ satisfies \eqref{block L}, with $\mathcal{L}^\mathcal{O}$ diagonal if $\Gamma^\mathcal{O}$ is. Also note that every digraph induced generator \eqref{generic} is pair block diagonal with respect to the standard basis with $\Gamma^\mathcal{O}$ diagonal and $\Gamma^\mathcal{D}=0$.

As noted before, $\gamma_{ij}\geq0$ since these are diagonal entries of positive semidefinite $\Gamma^\mathcal{O}$. It is not true in general, however, that $\gamma_{ii}\geq 0$, or that $\gamma_{ii}$ is even real. Indeed, considering the simple case of $\Gamma=\left(\begin{smallmatrix}0 & 0 \\ 0 & \Gamma^{\mathcal{D}}
\end{smallmatrix}\right)$, the criteria of Theorem~\ref{GKSL2} are satisfied for both $\Gamma^\mathcal{D}=\left(\begin{smallmatrix}
-1 & 0 & 0 \\ 0 & 2 & 0\\ 0 & 0 & 2
\end{smallmatrix}\right)$ and $\Gamma^\mathcal{D}=\left(\begin{smallmatrix}
-\imath & -\imath & -\imath \\ \imath & \imath & \imath\\ 0 & 0 & 0
\end{smallmatrix}\right)$.

Some things can still be said in our case of interest, though, as $\Gamma=\left(\begin{smallmatrix}
\Gamma^{\mathcal{O}} & 0 \\ 0 & \Gamma^\mathcal{D}
\end{smallmatrix}\right)$ satisfies the conditions of Theorem~\ref{GKSL2} if and only if $\Gamma=\left(\begin{smallmatrix}
\Gamma^{\mathcal{O}} & 0 \\ 0 & 0
\end{smallmatrix}\right)$ and $\Gamma=\left(\begin{smallmatrix}
0 & 0 \\ 0 & \Gamma^\mathcal{D}
\end{smallmatrix}\right)$ do. In particular, since $E_{ii}-E_{jj}$ is traceless it follows that \begin{align*}\langle E_{ii}-E_{kk},\Gamma^\mathcal{D}(E_{ii}-E_{jj})\rangle &= \Tr\left((E_{ii}-E_{jj})\left(\sum_{k=1}^N(\gamma_{ki}-\gamma_{kj})E_{kk}\right)\right) \\ &= \gamma_{ii}+\gamma_{jj}-\gamma_{iijj}-\gamma_{jjii}\geq 0. \end{align*}

We will recall this later as the following:

\begin{remark} \label{pos sums} If $\Gamma=\left(\begin{smallmatrix}
	\Gamma^{\mathcal{O}} & 0 \\ 0 & \Gamma^\mathcal{D}
	\end{smallmatrix}\right)$ then $\gamma_{ii}+\gamma_{jj}-\gamma_{iijj}-\gamma_{jjii}\geq0$ for all $1\leq i,j\leq N$. \end{remark}

\subsection{The Gell-Mann Basis}\label{GM basis}

By the Gell-Mann basis we mean the collection consisting of the normalized $N\times N$ identity matrix $\frac{1}{\sqrt{N}}I_N$ and three other sets of matrices: \begin{enumerate}
	\item[1)] The $\frac{N(N-1)}{2}$ many symmetric matrices defined by $$\lambda_{ij}:=\frac{1}{\sqrt{2}}(E_{ij}+E_{ji})\qquad \textrm{ for } 1\leq i < j \leq N,$$
	\item[2)] the $\frac{N(N-1)}{2}$ many antisymmetric matrices defined by $$\lambda_{ji}:=\frac{-\imath}{\sqrt{2}}(E_{ij}-E_{ji})\qquad \textrm{ for }1\leq i < j \leq N,$$
	\item[3)] and the $N-1$ many diagonal matrices defined by  $$\lambda_{nn}:=\frac{1}{\sqrt{n(n+1)}}\left(\sum_{m=1}^n E_{mm}-n E_{n+1,n+1}\right)\qquad \textrm{ for }1\leq n \leq N-1.$$
\end{enumerate} Each $\lambda_{ij}$ is Hermitian and traceless by construction, and they are orthonormal and orthogonal to $\frac{1}{\sqrt{N}}I_N$ in the Hilbert-Schmidt inner product \cite{BK}. By dimension count, we see that $\Span(\lambda_{ij},\frac{1}{\sqrt{N}}I_N)=M_N(\C)$.

Given a matrix written in the Gell-Mann basis, it is immediate how to write it in the standard basis. For the opposite direction, we use the formula given in \cite{BK}: \begin{equation} \label{EtoGM} E_{ij}=\left\{\begin{array}{ll}
\frac{1}{\sqrt{2}}(\lambda_{ij}+\imath\lambda_{ji}) & \textrm{ for } i<j \\
\frac{1}{\sqrt{2}}(\lambda_{ij}-\imath\lambda_{ji}) & \textrm{ for } j<i \\
-\sqrt{\frac{j-1}{j}}\lambda_{j-1,j-1}+\displaystyle\sum_{m=j}^{N-1}\textstyle\frac{1}{\sqrt{m(m+1)}}\lambda_{mm}+\frac{1}{N}I_N & \textrm{ for } i=j 
\end{array} 
\right.\end{equation} where the summation is interpreted as vacuously zero for $j=N$ and we take $\lambda_{00}:=0$.

Since the Gell-Mann basis without $I_N/\sqrt{N}$ is a complete set of traceless orthonormal matrices, given any QDS $T_t$ we may use Theorem~\ref{GKSL} to write its generator $\mathcal{L}$ with respect to the Gell-Mann basis: \begin{equation} \label{GM form} \mathcal{L}(\rho)=-\imath[H,\rho]+\frac{1}{2}\sum c_{ijk\ell}\left(\left[\lambda_{ij},\rho \lambda_{k\ell}\right]+\left[\lambda_{ij}\rho,\lambda_{k\ell}\right]\right)\end{equation} Note that no adjoints appear since each $\lambda_{ij}$ is Hermitian, and the sum is over all valid choices of $i,j,k ,\ell$; specifically, $i,j\in\{1,\ldots,N\}$ for $i\neq j$ and $i,j\in\{1,\ldots,N-1\}$ for $i=j$, and similarly  $k,\ell\in\{1,\ldots,N\}$ for $k\neq \ell$ and $k,\ell\in\{1,\ldots,N-1\}$ for $k=\ell$. We henceforth reserve $C$ to denote the $(N^2-1)\times (N^2-1)$ coefficient matrix $C:=(c_{ijk\ell})$ for $\mathcal{L}$ written with respect to the Gell-Mann basis, and \[D_{ijk\ell}^\lambda:=[\lambda_{ij},\cdot\lambda_{k\ell}]+[\lambda_{ij}\cdot,\lambda_{k\ell}]\] to denote the individual {\bf Gell-Mann basis Lindblad operators}.

Order the Gell-Mann basis as we did the standard basis, by pairing together $\lambda_{ij}$ and $\lambda_{ji}$ for $i\neq j$, then adjoining the diagonal $\lambda_{nn}$, and finally $I_N/\sqrt{N}$. Define $C^{\mathcal{O}}$ and $C^{\mathcal{D}_0}$ analogously as well, where now $\mathcal{D}_0:=\Span(\lambda_{ii})_{i=1}^{N-1}$ is the {\bf traceless diagonal subspace} of $M_N(\C)$, so that  $C^{\mathcal{D}_0}:\mathcal{D}_0\to \mathcal{D}_0$ is an $(N-1)\times (N-1)$ matrix. We use $a_{ij},b_{ij}$ and $c_{ij}$ for entries of $C$ as we used the notations $\alpha_{ij},\beta_{ij}$ and $\gamma_{ij}$ for entries of $\Gamma$.

To illustrate these notations, the following is an example of a matrix $C$ in dimension $N=3$ for which $C^{\mathcal{O}}$ is pair block diagonal and $C^{\mathcal{D}_0}$ is diagonal: $$C=\left(\begin{smallmatrix}
c_{12} & a_{12}+\imath b_{12} &  &  &  &  &  & \\ 
a_{12}-\imath b_{12} & c_{21} &  &  &  &  &  & \\ 
&  & c_{13} & a_{13}+\imath b_{13} &  &  &  & \\ 
&  & a_{13}-\imath b_{13} & c_{31} &  &  &  & \\ 
&  &  &  & c_{23} & \alpha_{23}+\imath b_{23} &  & \\ 
&  &  &  & a_{23}-\imath b_{23} & c_{32} &  & \\ 
&  &  &  &  &  & c_{11} & \\ 
&  &  &  &  &  &  & c_{22}\\
\end{smallmatrix}\right)$$

Motivated by the distinction between $\mathcal{D}$ and $\mathcal{D}_0$, let us denote by $\mathcal{L}^{\mathcal{D}_0}$ the submatrix of $\mathcal{L}$ formed by the rows and columns corresponding to diagonal $\lambda_{nn}$ for $1\leq n\leq N-1$. Explicitly, $$\mathcal{L}^{\mathcal{D}}=\begin{pmatrix}
\mathcal{L}^{\mathcal{D}_0} & \ast  \\ 
0 & 0
\end{pmatrix},$$ where the last row is zero since $\mathcal{L}$ has traceless range.

Under certain restrictions the matrix representations for $C$ and $\mathcal{L}$ with respect to the Gell-Mann basis \eqref{GM form} are unsurprisingly similar to those of $\Gamma$ and $\mathcal{L}$ with respect to the standard basis \eqref{E form}. Indeed, consider the basis change from the standard basis to the Gell-Mann basis represented by unitary matrix $U$, so that $\Gamma=U^\ast \widetilde{C}U$, where $\widetilde{C}$ is the matrix $C$ extended to act on all of $M_N(\C)$ by setting $\widetilde{C}(I_N)=0$ (i.e., $\widetilde{C}=\left(\begin{smallmatrix} C & 0 \\ 0 & 0\end{smallmatrix}\right)$). Then \eqref{EtoGM} implies $U=\left(\begin{smallmatrix}
U^\mathcal{O}& 0 \\ 0 & U^\mathcal{D}\end{smallmatrix}\right)$ where $ U^\mathcal{O}$ is pair block diagonal with each $ij$ block given by $\frac{1}{\sqrt{2}}\left(\begin{smallmatrix} 1 & 1 \\ \imath & -\imath \end{smallmatrix}\right)$ by \eqref{EtoGM}. We have general $ij$ blocks of the two forms are related via \begin{equation} \label{c to gamma} \begin{aligned}
\begin{pmatrix}
c_{ij} & a_{ij}+\imath b_{ij} \\ 
a_{ij}-\imath b_{ij} & c_{ji}
\end{pmatrix}^C&\equiv\frac{1}{2}\begin{pmatrix}
c_{ij}+c_{ji}-2b_{ij} & c_{ij}-c_{ji}-2\imath a_{ij} \\ 
c_{ij}-c_{ji}+2\imath a_{ij} & c_{ij}+c_{ji}+2b_{ij}
\end{pmatrix}^\Gamma \\ \begin{pmatrix}
\gamma_{ij} & \alpha_{ij}+\imath \beta_{ij} \\ 
\alpha_{ij}-\imath \beta_{ij} & \gamma_{ji}
\end{pmatrix}^\Gamma&\equiv\frac{1}{2}\begin{pmatrix}
\gamma_{ij}+\gamma_{ji}+2\alpha_{ij} & -2\beta_{ij}-\imath(\gamma_{ij}-\gamma_{ji}) \\ 
-2\beta_{ij}+\imath(\gamma_{ij}-\gamma_{ji}) & \gamma_{ij}+\gamma_{ji}-2\alpha_{ij}
\end{pmatrix}^C\end{aligned},\end{equation} where $\equiv$ denotes equal contribution to $\mathcal{L}$. This shows that for every $C=\left(\begin{smallmatrix} C^\mathcal{O} & 0 \\ 0 & C^{\mathcal{D}_0}\end{smallmatrix}\right)$ with $C^{\mathcal{O}}$ pair block diagonal there is some $\Gamma=\left(\begin{smallmatrix} \Gamma^\mathcal{O} & 0 \\ 0 & \Gamma^\mathcal{D}\end{smallmatrix}\right)$ with $\Gamma^{\mathcal{O}}$ pair block diagonal such that $C\equiv \Gamma$ (and vice-versa, up to Hamiltonian). Thus, assuming $H=\sum_{n=1}^Nh_nE_{nn}$, so that for $k<\ell$ we have \[-\imath[H,\lambda_{k\ell}]=\frac{-\imath}{\sqrt{2}}\sum_{n=1}^Nh_n[E_{nn},E_{k\ell}+E_{\ell k}]=(h_k-h_\ell)\lambda_{\ell k}\] and similarly $-\imath[H,\lambda_{\ell k}]=-(h_k-h_\ell)\lambda_{k\ell}$, from Remark~\ref{pair block remark} we have the following:

\begin{remark} \label{block L remark} Let $\mathcal{L}$ be a QDS generator written with respect to the Gell-Mann basis \eqref{GM form} with Hamiltonian $H=\sum_{n=1}^Nh_nE_{nn}$. If $C=\left(\begin{smallmatrix} C^\mathcal{O} & 0 \\ 0 & C^{\mathcal{D}_0}\end{smallmatrix}\right)$ with $C^{\mathcal{O}}$ pair block diagonal then \begin{equation} \label{block L2} \mathcal{L}=\begin{pmatrix}
	\mathcal{L}^{\mathcal{O}} & 0 & 0 \\ 
	0 & \mathcal{L}^{\mathcal{D}_0} & \ast \\ 
	0 & 0 & 0
	\end{pmatrix}=\begin{pmatrix}
	\mathcal{L}^{\mathcal{O}} & 0 \\ 
	0 & \mathcal{L}^{\mathcal{D}} 
	\end{pmatrix}, \end{equation} with $\mathcal{L}^{\mathcal{O}}$ pair block diagonal; in this case, if $C^{\mathcal{O}}$ is diagonal and $H=0$ then $\mathcal{L}^{\mathcal{O}}$ diagonal.
\end{remark}

A partial converse is also true, in the sense that no entry of $H$ outside its diagonal and no entry of $C$ outside both $C^{\mathcal{D}_0}$ and the pair block diagonal of $C^\mathcal{O}$ contributes to the pair block diagonal of $\mathcal{L}^{\mathcal{O}}$, to $\mathcal{L}^{\mathcal{D}_0}$, or to the portion of the $\mathcal{L}$ marked by $\ast$ in \eqref{block L2}. We also note that if $C^{\mathcal{O}}$ is diagonal and $C^{\mathcal{D}_0}$ is arbitrary then $\mathcal{L}(I_N)=0$ (and hence $\ast=0$) is easily verified.

\begin{definition} \label{convention2}
	We call QDS generator $\mathcal{L}$ {\bf pair block diagonal with respect to the Gell-Mann basis} if $\mathcal{L}$ is of form \eqref{GM form} with \begin{equation*} C=\begin{pmatrix} C^{\mathcal{O}} & 0 \\ 0 & C^{\mathcal{D}_0}\end{pmatrix}\end{equation*} and $C^{\mathcal{O}}$ pair block diagonal.
\end{definition}

Note that a QDS generator can be written as pair block diagonal with respect to the Gell-Mann basis if and only if it can be written as pair block diagonal with respect to the standard basis.

For basis-free definitions one may define $\mathcal{L}^\mathcal{D}:=P_{\mathcal{D}}\mathcal{L}|_\mathcal{D}$, where $P_\mathcal{D}$ is orthogonal projection onto $\mathcal{D}$, and similarly $\mathcal{L}^{\mathcal{D}_0}:=P_{\mathcal{D}_0}\mathcal{L}|_{\mathcal{D}_0}$. In the case $\mathcal{L}$ is of the form \eqref{block L2}, it follows from \eqref{kerstates} that $\ker\mathcal{L}^\mathcal{D}$ is nonempty, spanned by diagonal states (i.e., diagonal as $N\times N$ matrices), and it is natural to view $\ker\mathcal{L}^{\mathcal{D}_0}\subseteq\ker\mathcal{L}^{\mathcal{D}}$. It turns out this is true for arbitrary generators.

\begin{proposition}\label{DI=D+1}
	Let $\mathcal{L}$ be a QDS generator. Then $\ker\mathcal{L}^\mathcal{D}$ is nonempty, spanned by diagonal states, and  \[\ker\mathcal{L}^{\mathcal{D}}=\ker\mathcal{L}^{\mathcal{D}_0}\oplus\C\{\rho\}\] for any $\rho\in\ker\mathcal{L}^{\mathcal{D}}$ with nonzero trace. In particular, $\dim\ker\mathcal{L}^{\mathcal{D}}=\dim\ker\mathcal{L}^{\mathcal{D}_0}+1$.
\end{proposition}

\begin{proof}
	Without loss of generality assume $\mathcal{L}$ is written in Gell-Mann form \eqref{GM form}, and consider the matrix $\ol C$ obtained by setting equal to zero all entries of $C$ except those in the pair block diagonal of $C^{\mathcal{O}}$. Then the operator $\ol{\mathcal{L}}$ defined via \eqref{GM form} (with $H=0$) is a QDS generator, since $\ol C$ is positive semidefinite as each $ij$ block of $C$ is. Further, Remark~\ref{block L remark} and the partial converse thereof imply $\ol{\mathcal{L}}^\mathcal{D}=\mathcal{L}^{\mathcal{D}}$, and so we may assume without loss of generality that $C=\ol{C}$. From \eqref{kerstates} we conclude $\ker\mathcal{L}$ is nonempty and spanned by states. The block form \eqref{block L2} of $\mathcal{L}$ then implies $\ker\mathcal{L}^\mathcal{D}$ is nonempty and spanned by diagonal states. We now only need remark that given diagonal states $\rho_1,\rho_2\in\ker\mathcal{L}^\mathcal{D}$ we have that $\rho_1-\rho_2$ is diagonal, traceless, and in $\ker\mathcal{L}$, and hence $\rho_1-\rho_2\in\ker\mathcal{L}^{\mathcal{D}_0}$; that is, given fixed diagonal state $\rho_0\in\ker\mathcal{L}^{\mathcal{D}}$ we have that for any diagonal state $\rho\in\ker\mathcal{L}^{\mathcal{D}}$ there exists some diagonal traceless $A\in\ker\mathcal{L}^{\mathcal{D}_0}$ such that $\rho=\rho_0+A$. The dimensionality statement follows since every element in $\ker\mathcal{L}^{\mathcal{D}_0}$ is traceless but $\rho_0\in\ker\mathcal{L}^{\mathcal{D}}$ has unit trace.
\end{proof}
	
\section{Graph Theory Background}

In this section we establish notation and background for the needed graph theoretical notions; see \cite{chung} or any comparable text on elementary graph theory.

\subsection{Graphs}\label{graphsection}
	A {\bf graph} consists of a set of vertices, labeled $1,\ldots, N$, together with a set of weighted edges, which are 2-element sets $ij:=\{i,j\}$ of vertices each with an associated weight $w_{ij}>0$. A graph is called {\bf connected} if there is a path between every pair of vertices, and called a {\bf tree} if there is a unique path between every pair of vertices. Each maximal connected subgraph is called a {\bf connected component}. If $G$ is a graph on $N$ vertices, by its {\bf graph Laplacian} $L(G)$ we mean the $N\times N$ matrix whose $(i,j)$ entry is given by \[(L(G))_{ij}=\left\{\begin{array}{ll}
	 w_{ij} & i\neq j \\
	-\sum_{k\neq j}w_{kj} & i=j
	\end{array} 
	\right.,\] where we take $w_{ij}=0$ if $ij$ is not an edge of $G$.
	
	It is easy to see that $x^*L(G)x=\frac{-1}{2}\sum_{i,j=1}^Nw_{ij}|x_i-x_j|^2\leq 0$ for all vectors $x\in \mathbb{C}^N$, and so $L(G)$ is negative semidefinte. Notice that this quadratic form is zero if and only if $w_{ij}=0$ whenever $x_i\neq x_j$. Hence, if $G$ is connected the only vectors satisfying $x^\ast L(G)x=0$ are multiples of $\vec{1}$, the all ones vector, and so $\ker L(G)=\C\vec{1}$. If $G$ is not connected, then given connected components $G^1,\ldots,G^k$ of $G$ one may permute the underlying basis so that $L(G)$ is block diagonal of the form \[L(G)=\begin{pmatrix}
	L(G^1) & 0 & \cdots & 0 \\ 
	0 & L(G^2) & \cdots & 0 \\ 
	\vdots & \vdots & \ddots & \vdots \\ 
	0 & 0 & \cdots & L(G^k)
	\end{pmatrix},\] from which we establish the following well-known fact:\begin{remark} \label{gamma basis} For each connected component $G^n$ of a graph $G$ let $\gamma^{G^n}$ be the vector with one at each entry corresponding to a vertex in $G^n$ and zero elsewhere. Then $\Span(\gamma^{G^n})_{n=1}^k= \ker(L(G))$.
	\end{remark}

	\subsection{Digraphs}\label{digraphsection}	
	 A {\bf digraph} $G$ consists of a set $V(G)$ of vertices, labeled $1,\ldots, N$, together with a set $E(G)$ of weighted edges, which are ordered pairs $ij:=(i,j)$ of vertices each with an associated weight $w_{ji}>0$ (note the reversal of the indices). We regard edges $ij$ as the arrow from vertex $i$ to vertex $j$. A digraph is called a {\bf directed tree} if the graph obtained by ignoring the directedness of the edges is a tree. The {\bf weight of a directed tree} $T$ is is given by $\prod_{k\ell\in E(T)}w_{\ell k}$. We say $T$ is a {\bf directed spanning subtree} if $T$ is a subdigraph of $G$ which is a directed tree and $V(T)=V(G)$; we say further that $T$ is {\bf rooted at} $\boldsymbol{i}\in V(T)$ if $i$ is the only vertex of $T$ with no out-edges (in $T$). Denote by $\mathcal{T}_i(G)$ the collection of all directed spanning subtrees of $G$ rooted at $i$. If $G$ is a digraph on $N$ vertices, by {\bf digraph Laplacian} $L(G)$ we mean the $N\times N$ matrix whose $(i,j)$ entry is given by \[(L(G))_{ij}=\left\{\begin{array}{ll}
	w_{ij} & i\neq j \\
	-\sum_{k\neq j}w_{kj} & i=j
	\end{array} 
	\right.,\] where we take $w_{ji}=0$ if $ij$ is not an edge of $G$. By $L_k(G)$ we mean the $(N-1)\times (N-1)$ matrix obtained by deleting row $k$ and column $k$ from $L(G)$. 
	
	\begin{theorem}[\cite{tutte}]
		Let $G$ be a weighted digraph on $N$ vertices and let $L(G)$ be the corresponding digraph Laplacian. Then the total weight of all directed spanning subtrees of $G$ rooted at $i$ is given by \[\sum_{T\in\mathcal{T}_i(G)}\prod_{k\ell\in E(T)}w_{\ell k}=(-1)^{N-1}\det(L_i(G)).\]
	\end{theorem}
	
	 A digraph is called {\bf strongly connected} if between any two distinct vertices $i$ and $j$ there is a path from $i$ to $j$ and a path from $j$ to $i$. Each maximal strongly connected subdigraph is called a {\bf strongly connected component (SCC)}. Following Mirzaev and Gunawardena in \cite{MG}, we denote the SCC containing vertex $i$ as $[i]$, and write $[i]\preceq[j]$ if there is a path from $i'$ to $j'$ for some $i'\in[i]$ and $j'\in[j]$. If $[i]\preceq[j]$ implies $[i]=[j]$ for any $[j]$, we say $[i]$ is a {\bf terminal SCC (TSCC)}.
	 
	 For each TSCC $G^n$ of $G$ define vector $\widetilde{\rho}^{G_n}\in \R^N$ (where $N=|V(G)|$) by setting $\widetilde{\rho}_i^{G^n}$ to be the total weight of directed spanning subtrees of $G^n$ rooted at $i$; that is, \[\widetilde{\rho}_i^{G^n}=\sum_{T\in \mathcal{T}_i(G^n)}\prod_{k\ell\in E(T)}w_{\ell k}=(-1)^{N-1}\det(L_i(G^n)),\] where this quantity is taken to be zero if $i\not\in G^n$. We define \[\rho^{G^n}=\frac{1}{\lambda}\widetilde{\rho}^{G^n},\] where the normalization factor $\lambda>0$ is chosen so that $\sum_{i=1}^N\rho_i^{G^n}=1$ (explicitly, \linebreak$\lambda=(-1)^{N-1}\sum_i\det(L_i(G^n))$).
	 
	 \begin{proposition}[\cite{MG}] \label{rho basis} Let $G$ be a digraph (with all positive weights). Then \[\ker L(G)=\Span(\rho^{G^n})_{n=1}^k,\] where $G^1,\ldots,G^k$ are the TSCCs of $G$.
	 \end{proposition}
	 
	 By a {\bf sink} of a digraph we mean a single vertex which forms a TSCC; i.e., a vertex from which no edges originate. In a similar fashion, we call a pair of vertices $k$ and $\ell$ a {\bf 2-sink} if they form a TSCC; that is, there is an edge from $k$ to $\ell$ and vice versa, but no other edges originate from $k$ or $\ell$. If the context is clear, we denote a 2-sink on vertices $k$ and $\ell$ simply by the edge notation $k\ell$.
	 
\section{Relating Generators to Digraphs}
\subsection{Generator Induced Digraphs}\label{maindigraphsection}
Given a QDS generator $\mathcal{L}$, we define our main digraph of interest $G_\mathcal{L}$ to be the weighted digraph on $N$ vertices (labeled $1,2,\ldots, N$) with weight of edge from $j$ to $i$ (with $i\neq j$) given by $\gamma_{ij}$, where $\gamma_{ij}$ are the (uniquely determined by Theorem~\ref{GKSL2}) entries of $\Gamma^\mathcal{O}$ when $\mathcal{L}$ is written with respect to standard basis \eqref{E form}. Equivalently, \eqref{c to gamma} reveals that one may write $\mathcal{L}$ with respect to the Gell-Mann basis \eqref{GM form} and define  $G_\mathcal{L}$ to be the weighted digraph on $N$ vertices (labeled $1,2,\ldots, N$) with weight of edge from $j$ to $i$ given by \[\gamma_{ij}=\frac{1}{2}\left\{
\begin{array}{ll} c_{ij}+c_{ji}-2b_{ij} & i<j \\
c_{ji}+c_{ij}+2b_{ji} & i>j 
\end{array} 
\right. .\]  We note that \begin{equation}\label{smallrealpart}\frac{c_{ij}+c_{ji}}{2}\geq \sqrt{c_{ij}c_{ij}} \geq \sqrt{a_{ij}^2+b_{ij}^2}\geq|b_{ij}|,\end{equation} where the first inequality is a comparison of arithmetic and geometric means, and the second follows since the $ij$ block of $C$ is positive semidefinite (as $C$ itself is). Further, these inequalities are equality only in the case $c_{ij}=c_{ji}=|b_{ij}|$ and $a_{ij}=0$. Hence the following: \begin{remark} \label{positive G} The weights of graph $G_\mathcal{L}$ are nonnegative. Fix $i<j$. Then $\gamma_{ij}=0$ if and only if the $ij$ block of $C$ is given by $c_{ij}\begin{pmatrix}
	1 & \imath \\ 
	-\imath & 1
	\end{pmatrix}$, and $\gamma_{ji}=0$ if and only if the $ij$ block of $C$ is given by $c_{ij}\begin{pmatrix}
	1 & -\imath \\ 
	\imath & 1
	\end{pmatrix}$.\end{remark}

The following proposition shows that every QDS is naturally associated to a digraph.

\begin{theorem} \label{LD} Let $\mathcal{L}$ be a QDS generator written in matrix form with respect to the standard basis \eqref{matrix L}. Then $\mathcal{L}^{\mathcal{D}}=L(G_\mathcal{L})$. \end{theorem}

\begin{proof} Consider $\mathcal{L}$ given by form \eqref{E form}. The Hamiltonian part $i[H,\cdot]$ does not contribute to $\mathcal{L}^{\mathcal{D}}$ since evaluating $[H,E_{nn}]$ yields a matrix with null diagonal (explicitly, the $n$th column of $H$ minus the $n$th row of $H$). To find the contribution of the dissipative part, from \eqref{E calculation} we find
\begin{equation*}
D_{ijk\ell}(E_{nn}) =2\delta_{jn}\delta_{\ell n}E_{ik}-\delta_{\ell n}\delta_{ik}E_{nj}-\delta_{ik}\delta_{jn}E_{\ell n}.\end{equation*} Hence, $D_{ijk\ell}(E_{nn})$ has diagonal output if and only if $j=\ell=n$ and $i=k$, in which case $D_{ijij}(E_{jj})=2E_{ii}-2E_{jj}$. We have that $\mathcal{L}(E_{jj})$ has diagonal given by $\sum_{i\neq j}\gamma_{ij}(E_{ii}-E_{jj})$, and thus $\mathcal{L}^{\mathcal{D}}$ is given by \[(\mathcal{L}^{\mathcal{D}})_{iijj}=\left\{\begin{array}{ll}
\gamma_{ij} & i\neq j \\
-\sum_{k\neq j}\gamma_{kj} & i=j
\end{array} 
\right..\] 
\end{proof}

\begin{remark} \label{LD hermitian}
	If $G_\mathcal{L}$ satisfies $\gamma_{ij}=\gamma_{ji}$ for all pairs $i,j$, then $\mathcal{L}^\mathcal{D}$ is negative semidefinite (since undirected graph Laplacians are always negative semidefinite, as shown in Section~\ref{graphsection}).
\end{remark}

Recall Proposition~\ref{rho basis}, which states that vectors $\rho^{G_\mathcal{L}^n}$ give rise to a natural basis of $\ker L(G_\mathcal{L})$. Considering TSCCs $G_\mathcal{L}^1,\ldots,G_\mathcal{L}^k$ of $G_\mathcal{L}$, we write these vectors as matrices by defining  \begin{equation} \label{d_n} d^{G^n_\mathcal{L}}:=\sum_{i=1}^N\rho_{i}^{G_\mathcal{L}^n}E_{ii}=\sum_{i=1}^{N-1}\left(\sum_{j=1}^{i}\rho_{j}^{G_\mathcal{L}^n}-i\rho_{i+1}^{G_\mathcal{L}^n}\right)\frac{\lambda_{ii}}{\sqrt{i(i+1)}}+\frac{I_N}{N}\qquad 1\leq n\leq k,\end{equation} where the second equality can be checked using \eqref{EtoGM}. From Proposition~\ref{LD} and Proposition~\ref{rho basis} follows the analogous result:

\begin{corollary} \label{general ker LDI} Let $\mathcal{L}$ be a QDS generator. Let $G_\mathcal{L}^1,\ldots,G_\mathcal{L}^k$ denote the TSCCs of $G_\mathcal{L}$. Then \[\ker\mathcal{L}^{\mathcal{D}}=\Span\left(d^{G^n_\mathcal{L}}\right)_{n=1}^k.\]
\end{corollary}

In the case $\gamma_{ij}=\gamma_{ji}$ for all pairs $i,j$ (for example, if $\mathcal{L}$ arises from diagonal $C$), then a basis for $\ker\mathcal{L}^{\mathcal{D}}$ is easier to compute. Indeed, considering the digraph $G_\mathcal{L}$ as an undirected graph $H_\mathcal{L}$, for each connected component $H_\mathcal{L}^1,\ldots,H_\mathcal{L}^k$ of $H_\mathcal{L}$ we may use the simpler vectors $\gamma^{H_\mathcal{L}^n}$ given in Remark~\ref{gamma basis} to define \begin{equation} \label{tilded_n} d^{H^n_\mathcal{L}} =\sum_{i=1}^N\gamma_{i}^{H_\mathcal{L}^n}E_{ii}=\sum_{i=1}^{N-1}\left(\sum_{j=1}^{i}\gamma_{j}^{H_\mathcal{L}^n}-i\gamma_{i+1}^{H_\mathcal{L}^n}\right)\frac{\lambda_{ii}}{\sqrt{i(i+1)}}\qquad 1\leq n\leq k, \end{equation} and establish the following result:

\begin{proposition} \label{ker LDI} Let $\mathcal{L}$ be a QDS generator such that $\gamma_{ij}=\gamma_{ji}$ for all pairs $i\neq j$. Let $H_\mathcal{L}^1,\ldots,H_\mathcal{L}^k$ denote the connected components of $H_\mathcal{L}$. Then \begin{align*}\ker\mathcal{L}^{\mathcal{D}} &= \Span(d^{H^n_\mathcal{L}})_{n=1}^k.\end{align*}
\end{proposition}

\section{Pair Block Diagonal $\mathcal{L}$}

\subsection{The $\boldsymbol{\mathcal{L}^{\mathcal{O}}}$ part of $\boldsymbol{\mathcal{L}}$}\label{section LO}

The previous section revealed that $\ker\mathcal{L}^{\mathcal{D}}$ is characterized by the TSCCs of $G_\mathcal{L}$. The aim of this section is to establish a similar result for $\mathcal{L}^{\mathcal{O}}$ when $\mathcal{L}$ is pair block diagonal. The type of TSCCs we require here is more precise, however, and we must begin by establishing a few definitions.

We call a 2-sink $k\ell$ of $G_\mathcal{L}$ a {\bf singular 2-sink} if $\gamma_{k\ell}=\gamma_{\ell k}$ and the $k\ell$ block of $\Gamma^{\mathcal{O}}$ is singular. Rephrased in terms of $C$, a 2-sink $k\ell$ of $G_\mathcal{L}$ is a singular 2-sink if $c_{k\ell}c_{\ell k}-a_{k\ell}^2=0$, as this equality implies $b_{k\ell}=0$ (equivalently $\gamma_{k\ell}=\gamma_{\ell k}$) by \eqref{smallrealpart}. We use $S_{G_\mathcal{L}}$ to denote the set of sinks of $G_\mathcal{L}$ and $S^2_{G_\mathcal{L}}$ to denote the set of singular 2-sinks of $G_\mathcal{L}$.

Notably, in the definition of singular 2-sinks we require information beyond the weights of $G_\mathcal{L}$, namely $\alpha_{k\ell}$ and $\beta_{k\ell}$. It follows that graph induced generators \eqref{generic} satisfy $S^2_{G_\mathcal{L}}=\emptyset$, as in this case the $k\ell$ block of $\Gamma^{\mathcal{O}}$ is always nonsingular unless it is identically zero, precluding the possibility of $k\ell$ to be a 2-sink. The next lemma shows further coefficients which are not graph induced, such as the entries of $\Gamma^{\mathcal{D}}$, also affect $\ker\mathcal{L}^{\mathcal{O}}$. Here we assume for simplicity that $\Gamma\geq0$ as in Theorem~\ref{GKSL3}, but we note after Theorem~\ref{eigenvalues E} how one may produce the statement for $\Gamma\not\geq 0$.

\begin{lemma} \label{LSA in E}
	Let $\mathcal{L}$ be a QDS generator which is pair block diagonal with respect to the standard basis \eqref{E form} with $\widetilde{H}=\sum_{n=1}^Nh_nE_{nn}$ and $\Gamma\geq 0$. Then the $k\ell$ block $\mathcal{L}_{k\ell}$ of $\mathcal{L}^\mathcal{O}$ is singular if and only if $h_k=h_\ell$, $\gamma_{kk}=\gamma_{\ell\ell}=\gamma_{kk\ell\ell}$, and either\begin{itemize}
		\item $k,\ell\in S_{G_\mathcal{L}}$, in which case $\ker\mathcal{L}_{k\ell}=\Span(E_{k\ell},E_{\ell k})$, or
		\item $k\ell\in S^2_{G_\mathcal{L}}$, in which case  $\ker\mathcal{L}_{k\ell}=\C\left\{(\gamma_{k\ell}+\alpha_{k\ell}+\imath\beta_{k\ell})E_{k\ell}+(\gamma_{k\ell}+\alpha_{k\ell}-\imath\beta_{k\ell})E_{\ell k}\right\}.$
	\end{itemize}
\end{lemma}

\begin{proof} We fix $k< \ell$ and calculate the exact matrix form of $\mathcal{L}_{k\ell}$ by evaluating $\mathcal{L}$ at $E_{k\ell}$ and $E_{\ell k}$. From \eqref{E calculation} we have $$\sum_{n,m=1}^{N}\gamma_{nnmm}D_{nnmm}(E_{k\ell}) = (2\gamma_{kk\ell\ell}-\gamma_{kk}-\gamma_{\ell\ell})E_{k\ell}$$ and $$\sum_{n,m=1}^{N}\gamma_{nnmm}D_{nnmm}(E_{\ell k}) = (2\gamma_{\ell\ell kk}-\gamma_{kk}-\gamma_{\ell\ell})E_{\ell k},$$  which is to say $\Gamma^{\mathcal{D}}$ contributes to $\mathcal{L}_{k\ell}$ the $2\times 2$ matrix \[D:=\frac{1}{2}\begin{pmatrix}
		2\gamma_{kk\ell\ell}-\gamma_{kk}-\gamma_{\ell\ell} & 0\\ 
		0 & 2\gamma_{\ell\ell kk}-\gamma_{kk}-\gamma_{\ell\ell}
	\end{pmatrix}=\begin{pmatrix}
	d_{k\ell} & 0\\ 
	0 & \ol{d}_{k\ell}
	\end{pmatrix},\] where we define $d_{k\ell}:=\gamma_{kk\ell\ell}-\frac{1}{2}(\gamma_{kk}+\gamma_{\ell\ell})$ for future notational convenience (and hence $\ol{d}_{k\ell}=\gamma_{\ell\ell kk}-\frac{1}{2}(\gamma_{kk}+\gamma_{\ell\ell})$ since $\Gamma\geq 0$). Remark~\ref{pos sums} gives that $\re d_{k\ell}\leq 0$, and so $D$ has eigenvalues in the closed right hand plane.
	
	Considering $\Gamma^{\mathcal{O}}$, from \eqref{E calculation} we have, for $i\neq j$,
	\[D_{ij}(E_{k\ell}) = -(\delta_{jk}+\delta_{j\ell})E_{k\ell}, \quad D_{ij}(E_{\ell k})=-(\delta_{j\ell}+\delta_{jk})E_{\ell k}\] and \[
	D_{ijji}(E_{k\ell}) = 2\delta_{jk}\delta_{i\ell}E_{\ell k},\quad D_{ijji}(E_{\ell k})=2\delta_{j\ell}\delta_{ik}E_{k\ell}.\] Thus, an $ij$ block of $\Gamma^{\mathcal{O}}$ for which $|\{i,j\}\cap\{k,\ell\}|=0$ contributes nothing to $\mathcal{L}_{k\ell}$, and an $ij$ block of $\Gamma^{\mathcal{O}}$ for which $|\{i,j\}\cap\{k,\ell\}|=1$ contributes to $\mathcal{L}_{k\ell}$ the $2\times 2$ matrix \[IJ:=\frac{1}{2}\left\{
	\begin{array}{ll} -\gamma_{ji}I_2 & i\in\{k,\ell\}\not\ni j \\
	-\gamma_{ij}I_2 &  i\not\in\{k,\ell\}\ni j 
	\end{array} 
	\right..\]
	Note that $IJ$ is negative semidefinite since $\gamma_{ij},\gamma_{ji}\geq 0$ (see Remark~\ref{positive G}). Note also that $IJ$ is singular if and only if $\gamma_{ji}=0$ when $i\in\{k,\ell\}$ or $\gamma_{ij}=0$ when $j\in\{k,\ell\}$, in which case $IJ=0$. 
	
	Similarly, the above equations show that the $k\ell$ block $\left(\begin{smallmatrix}
	\gamma_{k\ell} & \alpha_{k\ell}+\imath\beta_{k\ell} \\ \alpha_{k\ell}-\imath\beta_{k\ell} & \gamma_{\ell k}
	\end{smallmatrix}\right)$ of $\Gamma^{\mathcal{O}}$ contributes to $\mathcal{L}_{k\ell}$ the $2\times 2$ matrix \[KL:=\frac{1}{2}\begin{pmatrix}
	-\gamma_{k\ell}-\gamma_{\ell k} & 2(\alpha_{k\ell}+\imath\beta_{k\ell}) \\ 
	2(\alpha_{k\ell}-\imath\beta_{k\ell}) & -\gamma_{k\ell}-\gamma_{\ell k}
	\end{pmatrix}.\] Note that the $k\ell$ of $\Gamma^{\mathcal{O}}$ block is positive semidefinite, as it is a principal submatrix of positive semidefinite $\Gamma^{\mathcal{O}}$. Thus $KL$ is negative semidefinite, as it is the negated sum of the $k\ell$ block of $\Gamma^{\mathcal{O}}$ and its anti-diagonal transpose, both positive semidefinite matrices. Also note that $KL$ is singular if and only if $\det(KL)=0$.
	
	Finally, we compute \[-\imath[\widetilde{H},E_{k\ell}] = -\imath\sum_{n=1}^Nh_n[E_{nn},E_{k\ell}]=-\imath(h_k-h_\ell)E_{k\ell}\] and similarly $-\imath[\widetilde{H},E_{\ell k}]=-\imath(h_\ell-h_k)E_{\ell k}$, which is to say $\widetilde{H}$ contributes to $\mathcal{L}_{k\ell}$ the $2\times 2$ matrix \[\ol{H}:=\begin{pmatrix}
	-\imath(h_k-h_\ell) & 0\\ 
	0 & \imath(h_k-h_\ell)
	\end{pmatrix}.\]
	
	In total, we now have that \[\mathcal{L}_{k\ell}=KL+\ol{H}+D+\sum_{|\{i,j\}\cap\{k\ell\}|=1} IJ.\] We claim that $KL+\ol{H}+D$ has eigenvalues all in the closed left-hand plane. Indeed, if we consider the matrix $\widetilde{C}$ obtained by setting equal to zero all entries of $\Gamma$ except those in $\Gamma^\mathcal{D}$ and the $k\ell$ block of $\Gamma^\mathcal{O}$, then $\widetilde{\Gamma}\geq0$ and so $\widetilde{\mathcal{L}}$ is a QDS generator by Theorem~\ref{GKSL2}. Moreover, this has the affect of setting $IJ=0$ for all $IJ$ but leaving the other calculations unchanged above, and so we have $\widetilde{\mathcal{L}}_{k\ell}=KL+\ol{H}+D$. The block form \eqref{block L} implies every eigenvalue of $\widetilde{\mathcal{L}}_{k\ell}$ is an eigenvalue of $\widetilde{\mathcal{L}}$ and so must lie in the closed left-hand plane (if $\widetilde{\mathcal{L}}(x)=\lambda x$ then $\widetilde{T}_t(x)=e^{t\lambda}x$, and so $||T_t(x)||_1=|e^{t\lambda}|\Tr(|x|)\leq \Tr(|x|)=||x||_1$ implies $\re\lambda\leq 0$ since $||T_t||_{1\to1}\leq 1$ as remarked in Section~\ref{contractivitysection}). 

	Since $KL+\ol{H}+D$ and all $IJ$ pairwise commute (every $IJ$ is a multiple of $I_2$), every eigenvalue of $\mathcal{L}_{k\ell}$ is the sum of eigenvalues $KL+\ol{H}+D$ and each $IJ$. Since each $IJ$ is negative semidefinite and $KL+\ol{H}+D$ has eigenvalues in the closed left-hand plane, $\mathcal{L}_{k\ell}$ is singular (has eigenvalue 0) if and only if $KL+\ol{H}+D$ and each of the $IJ$ are singular; that is, $\mathcal{L}_{k\ell}$ is singular if and only if each of the following hold:\begin{itemize}
		\item[(i)\phantom{ii}] $\det(KL+\ol{H}+D)=0$
		\item[(ii)\phantom{i}] $\gamma_{ji}=0$ for all $i<j$ with $i\in\{k,\ell\}\not\ni j$
		\item[(iii)] $\gamma_{ij}=0$ for all $i<j$ with $i\not\in\{k,\ell\}\ni j$
	\end{itemize}

	We claim that condition (i) can be rewritten as \begin{itemize}\item[(i)]$ \gamma_{k\ell}=\gamma_{\ell k},\text{ the $k\ell$ block of $\Gamma^\mathcal{O}$ is singular},$ $\gamma_{kk}=\gamma_{\ell\ell}=\gamma_{kk\ell\ell}, \text { and }h_k=h_\ell$.\end{itemize} Indeed, using $d_{k\ell}=\gamma_{kk\ell\ell}-\frac{1}{2}(\gamma_{kk}+\gamma_{\ell\ell})$, $h_{k\ell}=h_k-h_\ell$, and $y_{k\ell}=\frac{1}{2}(\gamma_{k\ell}+\gamma_{\ell k})$ for notational convenience, we have $\det(KL+\ol{H}+D)=$ \begin{align*} &= (-y_{k \ell}+d_{k\ell}-\imath h_{k\ell})(-y_{k \ell}+\ol{d}_{k\ell}+\imath h_{k\ell})-(\alpha_{k\ell}+\imath\beta_{k\ell})(\alpha_{k\ell}-\imath\beta_{k\ell}) \\ &=  y_{k\ell}^2+(d_{k\ell}-\imath h_{k\ell})(\ol{d}_{k\ell}+\imath h_{k\ell})-y_{k\ell}(d_{k\ell}+\ol{d}_{k\ell})-\alpha_{k\ell}^2-\beta_{k\ell}^2.\end{align*} We understand this equation as three nonnegative parts:
	
	First, since the $k\ell$ block of $C$ is positive semidefinite, we have that \begin{align*}P_1:=y_{k\ell}^2-\alpha_{k\ell}^2-\beta_{k\ell}^2 &= (y_{k\ell}+\alpha_{ij})(y_{k\ell}-\alpha_{ij})-(-\beta_{ij})^2 \\ &=c_{k\ell}c_{\ell k}-a_{k\ell}^2\geq0 \end{align*} using conversion \eqref{c to gamma}. It follows that $P_1=0$ if and only if $\gamma_{k\ell}=\gamma_{\ell k}$ and the $k\ell$ block of $\Gamma^\mathcal{O}$ is singular, as remarked in the equivalent definitions of singular 2-sinks in the preamble of this section. 

Second,  \begin{align*} P_2:=(d_{k\ell}-\imath h_{k\ell})(\ol{d}_{k\ell}+\imath h_{k\ell}) &= (d_{k\ell}-\imath h_{k\ell})\ol{(d_{k\ell}-\imath h_{k\ell})}\geq0.\end{align*} Since $\Gamma$ is positive semidefinite the submatrix $\left(\begin{smallmatrix} \gamma_{kk} & \gamma_{kk\ell \ell} \\ \gamma_{\ell\ell kk} & \gamma_{\ell \ell}	\end{smallmatrix}\right)$ is as well, from which it follows that $$\textstyle-2\re(d_{k\ell}-\imath h_{k\ell})=-2\re(d_{k\ell})=-(d_{k\ell}+\ol{d}_{k\ell})=\gamma_{kk}+\gamma_{\ell\ell}-2\re(\gamma_{kk\ell \ell})\geq 0,$$ with equality if and only if $\gamma_{kk}=\gamma_{\ell\ell}=\gamma_{kk\ell\ell}$ (this follows identically as \eqref{smallrealpart}). In particular, $\re(d_{k\ell})=0$ implies $\im(d_{k\ell})=0$, so we have that $P_2=0$ if and only if $\gamma_{kk}=\gamma_{\ell\ell}=\gamma_{kk\ell\ell}$ and $h_{k\ell}=0$.

Finally, $$P_3:=-y_{k\ell}(d_{k\ell}+\ol{d}_{k\ell})=\frac{1}{2}(\gamma_{k\ell}+\gamma_{\ell k})(\gamma_{kk}+\gamma_{\ell\ell}-2\re(\gamma_{kk\ell\ell}))\geq 0,$$ with $P_3=0$ if and only if $\gamma_{kk}=\gamma_{\ell\ell}=\gamma_{kk\ell\ell}$ or $\gamma_{k\ell}=\gamma_{\ell k}=0$, with similar reasoning as above.

Thus, we have that $\det(KL+\ol{H}+D)=P_1+P_2+P_3=0$ if and only if $P_1=P_2=P_3=0$. By the arguments above, this happens if and only if the rephrased (i) holds.

The next two conditions (ii) and (iii) simply say that vertices $k$ and $\ell$ have no out edges, except possibly to each other. Thus, if (i) holds, this means either $\gamma_{k\ell}=\gamma_{\ell k}\neq 0$ and $k\ell$ is a singular 2-sink of $G_\mathcal{L}$, or $\gamma_{k\ell}=\gamma_{\ell k}=0$ and $k$ and $\ell$ are sinks of $G_\mathcal{L}$.

It remains to note that if $\mathcal{L}_{k\ell}$ is singular, and hence (i), (ii), and (iii) hold, then $\mathcal{L}_{k\ell}=KL$, as $\ol{H}$, $D$, and all $IJ$ are necessarily zero. Thus, if $\mathcal{L}_{k\ell}$ is singular then \begin{equation}\label{diamond}\ker\mathcal{L}_{k\ell}=\left\{
\begin{array}{ll}
\textstyle\C\{(\gamma_{k\ell}+\alpha_{k\ell}+\imath\beta_{k\ell})E_{k\ell}+(\gamma_{k\ell}+\alpha_{k\ell}-\imath\beta_{k\ell})E_{\ell k}\} & \textrm{if } k\ell\in S^2_{G_\mathcal{L}} \\
\Span(E_{k\ell},E_{\ell k}) & \textrm{if } k,\ell\in S_{G_\mathcal{L}} \\
\end{array} 
\right.,\end{equation} as can either be directly verified or obtained as a corollary of Theorem~\ref{eigenvalues E} (see Remark~\ref{eig cor}).
 \end{proof}

\begin{corollary}\label{negative E}
	Let $\mathcal{L}$ be a Hamiltonian-free QDS generator which is pair block diagonal with respect to the standard basis \eqref{E form} with $\Gamma^\mathcal{D}$ diagonal. Then $\mathcal{L}^\mathcal{O}$ is negative semidefinite.
\end{corollary}

\begin{proof}
	Considering a $k\ell$ block $\mathcal{L}_{k\ell}$ of $\mathcal{L}$ computed as in the proof of Lemma~\ref{LSA in E}, we have $\mathcal{L}_{k\ell}=KL+D+\sum_{|\{i,j\}\cap\{k\ell\}|=1} IJ$. As before, $KL$ and each $IJ$ is negative semidefinite, so it suffices to show that $D$ is negative semidefinite if $\Gamma^\mathcal{D}$ diagonal. This is indeed the case, since $D=\frac{1}{2}\left(\begin{smallmatrix}
		-\gamma_{kk}-\gamma_{\ell\ell} & 0\\ 
		0 & -\gamma_{kk}-\gamma_{\ell\ell}
	\end{smallmatrix}\right)$ and $\gamma_{kk}+\gamma_{\ell\ell}\geq 0$ by Remark~\ref{pos sums}.
\end{proof}

\begin{theorem} \label{eigenvalues E} Let $\mathcal{L}$ be a QDS generator which is pair block diagonal with respect to the standard basis \eqref{E form} with $\widetilde{H}=\sum_{n=1}^Nh_nE_{nn}$. Then the $k\ell$ block $\mathcal{L}_{k\ell}$ of $\mathcal{L}^\mathcal{O}$ has eigenmatrices $A^\pm=$\small\begin{align*} &\left[\alpha_{k\ell}+\imath\beta_{k\ell}+\frac{\gamma_{kk\ell\ell}-\gamma_{\ell\ell kk}}{2}-\imath(h_k-h_\ell)\pm\sqrt{\alpha_{k\ell}^2+\beta_{k\ell}^2+\left(\frac{\gamma_{kk\ell\ell}-\gamma_{\ell\ell kk}}{2}-\imath(h_k-h_\ell)\right)^2}\right]E_{k\ell} \\ +&\left[\alpha_{k\ell}-\imath\beta_{k\ell}-\frac{\gamma_{kk\ell\ell}-\gamma_{\ell\ell kk}}{2}+\imath(h_k-h_\ell)\pm\sqrt{\alpha_{k\ell}^2+\beta_{k\ell}^2+\left(\frac{\gamma_{kk\ell\ell}-\gamma_{\ell\ell kk}}{2}-\imath(h_k-h_\ell)\right)^2}\right]E_{\ell k}\end{align*} corresponding to eigenvalues \begin{align*} \mu^{\pm}=&-\frac{1}{2}\left(\gamma_{k\ell}+\gamma_{\ell k}+\gamma_{kk}+\gamma_{\ell\ell}-\gamma_{kk\ell\ell}-\gamma_{\ell\ell kk} +\sum_{i\not\in\{k,\ell\}\ni j}\gamma_{ij}+\sum_{i\in\{k,\ell\}\not\ni j}\gamma_{ji}\right) \\&\pm\sqrt{\alpha_{k\ell}^2+\beta_{k\ell}^2+\left(\frac{\gamma_{kk\ell\ell}-\gamma_{\ell\ell kk}}{2}-\imath(h_k-h_\ell)\right)^2}.\end{align*} In particular, $E_{k\ell}$ and $E_{\ell k}$ are eigenmatrices of $\mathcal{L}^{\mathcal{O}}$ if and only if $\alpha_{k\ell}=\beta_{k\ell}=0$, in which case they have eigenvalues $\gamma_{kk\ell\ell}-\imath(h_k-h_\ell)-\mu$ and $\gamma_{\ell\ell kk}-\imath(h_k-h_\ell)-\mu$, respectively, where \[\mu=\frac{1}{2}\left(\gamma_{k\ell}+\gamma_{\ell k}+\gamma_{kk}+\gamma_{\ell\ell}+\sum_{i\not\in\{k,\ell\}\ni j}\gamma_{ij}+\sum_{i\in\{k,\ell\}\not\ni j}\gamma_{ji}\right).\]  \end{theorem}

\begin{proof}	
	It is well known that given a $2\times 2$ matrix $M = \left( \begin{smallmatrix}a & b\\ c & d\end{smallmatrix}\right)$ its eigenvectors are given by $\left( \begin{smallmatrix}\mu^\pm+b-d\\ \mu^\pm +c-a\end{smallmatrix}\right)$, where $\mu^\pm=\Tr(M)/2\pm(\Tr^2(M)/4-\det(M))^{1/2}$ are the corresponding eigenvalues, as can be verified by simply evaluating $M$ at the proposed eigenvectors. This fact applied to $KL+\ol{H}+D$ (as compute in the proof of Lemma~\ref{LSA in E}), along with the shift from adding $\sum IJ$ (multiple of $I_2$) immediately gives the above formula.
\end{proof}

\begin{remark}\label{eig cor}
If $k\ell\in S^2_{G_\mathcal{L}}$ then $\gamma_{k\ell}^2-\alpha_{k\ell}^2-\beta_{k\ell}^2=0$ since the $k\ell$ block of $\Gamma^{\mathcal{O}}$ is singular. Hence, $\gamma_{k\ell}=\gamma_{\ell k}=\sqrt{\alpha_{k\ell}^2+\beta_{k\ell}^2}$ in this case. If we further assume $h_k=h_\ell$ and $\gamma_{kk}=\gamma_{\ell\ell}=\gamma_{kk\ell\ell}$, then we have that  $A^+=(\gamma_{k\ell}+\alpha_{k\ell}+\imath\beta_{k\ell})E_{k\ell}+(\gamma_{k\ell}+\alpha_{k\ell}-\imath\beta_{k\ell})E_{\ell k}$ corresponding to $\mu^+=0$ generates $\ker\mathcal{L}_{k\ell}$, as given before in \eqref{diamond}.
\end{remark}

We note two facts: First, $\Gamma\geq 0$ was not assumed in Theorem~\ref{eigenvalues E}, as the calculations needed did not rely on this fact. Hence, one may set $\mu^\pm=0$ to write Lemma~\ref{LSA in E} without the $\Gamma\geq 0$ assumption. Second, Theorem~\ref{eigenvalues E} provides an explicit formula for $N^2-N$ of $\mathcal{L}$'s $N^2$ many eigenpairs, but since the digraph Laplacian $\mathcal{L}^\mathcal{D}$ is not diagonalizable in general the entire matrix $\mathcal{L}$ may not be diagonalizable. Finding the eigenvalues of a digraph Laplacian is historically difficult, but much work has been done on finding the spectral gap, as this controls the rate of convergence of $e^{tL}$. Though we do not explore such applications in this work, we note that, together with the eigenvalues given by Theorem~\ref{eigenvalues E}, the spectral gap of $\mathcal{L}^\mathcal{D}$ gives the rate of convergence for $T_t=e^{t\mathcal{L}}$. We refer the interested reader to the seminal work of Wu \cite{wu} for more on the eigenvalues of digraph Laplacians.

Having established results for the standard basis, we now consider the Gell-Mann basis. Certainly one may use \eqref{c to gamma} and the corresponding equivalence for converting $C^{\mathcal{D}_0}$ into $\Gamma^\mathcal{D}$ to translate Theorem~\ref{eigenvalues E} immediately into the corresponding general statement for the Gell-Mann basis. As we will only consider the Gell-Mann basis in specialized cases, we avoid writing this tedious conversion here and instead prove the needed statement directly.

\begin{lemma} \label{diag LSA}
	Let $\mathcal{L}$ be a QDS generator which is pair block diagonal with respect to the Gell-Mann basis \eqref{GM form} with $H=\sum_{n=1}^Nh_nE_{nn}$ and $C^{\mathcal{D}_0}$ diagonal. Then the $k\ell$ block $\mathcal{L}_{k\ell}$ of $\mathcal{L}$ is singular if and only if $h_k=h_\ell$, $c_{nn}=0$ for all $k-1\leq n\leq \ell-1$, and \begin{itemize}
		\item $k,\ell\in S_{G_\mathcal{L}}$, in which case $\ker\mathcal{L}_{k\ell}=\Span(\lambda_{k\ell},\lambda_{\ell k})$, or
		\item $k\ell\in S^2_{G_\mathcal{L}}$, in which case $\ker\mathcal{L}_{k\ell}=\C\{(c_{k\ell}+a_{k\ell})\lambda_{k\ell}+(c_{\ell k} +a_{k\ell})\lambda_{\ell k}\}$.
	\end{itemize}
\end{lemma}

\begin{proof} As in the proof of Lemma~\ref{LSA in E}, we calculate $\mathcal{L}_{k\ell}$ explicitly. Indeed, the only difference here is the contribution of $C^{\mathcal{D}_0}$, since the contribution of $H$ and $C^{\mathcal{O}}$ can be recovered from the formula for $\ol{H}$, $IJ$, and $KL$ calculated there. Using the same basis change as in the derivation of \eqref{c to gamma}, these matrices are represented in the Gell-Mann basis as

	\[\textstyle\ol{H}=\left(\begin{smallmatrix} 0 & h_\ell-h_k \\ h_k-h_\ell & 0 \end{smallmatrix}\right),\text{ }KL=\left(\begin{smallmatrix}
	-c_{\ell k} & a_{k\ell} \\ 
	a_{k\ell} & -c_{k\ell}
	\end{smallmatrix}\right),\text{ and }IJ=-\frac{1}{4}\left\{
	\begin{array}{ll} (c_{ij}+c_{ji}+2b_{ij})I_2 & i\in\{k,\ell\}\not\ni j \\
	(c_{ij}+c_{ji}-2b_{ij})I_2 &  i\not\in\{k,\ell\}\ni j 
	\end{array} 
	\right..\]
	
By Appendix~\ref{Diijj appen} we have \[D_{nn}^\lambda(\lambda_{k\ell})	= \left\{
\begin{array}{ll}
\frac{-n}{n+1}\lambda_{k\ell} & n=k-1 \\
\frac{-1}{n(n+1)}\lambda_{k\ell} & k\leq n\leq\ell-2 \\
\frac{-(n+1)}{n}\lambda_{k\ell} & n=\ell-1 \\
0 & \textrm{otherwise}
\end{array} 
\right.\;,\qquad D_{nn}^\lambda(\lambda_{\ell k})	=\left\{
\begin{array}{ll}
\frac{-n}{n+1}\lambda_{\ell k} & n=k-1 \\
\frac{-1}{n(n+1)}\lambda_{\ell k} & k\leq n\leq\ell-2 \\
\frac{-(n+1)}{n}\lambda_{\ell k} & n=\ell-1 \\
0 & \textrm{otherwise}
\end{array} 
\right..\]

Thus, \[\sum_{n=1}^{N-1}c_{nn}D_{nn}^\lambda(\lambda_{k\ell}) = -\left(\frac{k-1}{k}c_{k-1,k-1}+\sum_{m=k}^{\ell-2}\frac{1}{{m(m+1)}}c_{mm}+\frac{\ell}{(\ell-1)}c_{\ell-1,\ell-1}\right)\lambda_{k\ell},\] \[\sum_{n=1}^{N-1}c_{nn}D_{nn}^\lambda(\lambda_{\ell k}) = -\left(\frac{k-1}{k}c_{k-1,k-1}+\sum_{m=k}^{\ell-2}\frac{1}{{m(m+1)}}c_{mm}+\frac{\ell}{(\ell-1)}c_{\ell-1,\ell-1}\right)\lambda_{\ell k},\] which is to say $C^{\mathcal{D}_0}$ contributes to $\mathcal{L}_{k\ell}$ the $2\times 2$ matrix \[D^C:=-\frac{1}{2}\left(\frac{k-1}{k}c_{k-1,k-1}+\sum_{m=k}^{\ell-2}\frac{1}{{m(m+1)}}c_{mm}+\frac{\ell}{\ell-1}c_{\ell-1,\ell-1}\right)I_2.\] Note that $D^C$ is negative semidefinite (each $c_{nn}\geq0$ since $C\geq0$). Furthermore, $D^C$ is singular if and only if $c_{nn}=0$ for all $k-1\leq n\leq \ell-1$, in which case $D^C=0$.

In total, we now have that \[\mathcal{L}_{k\ell}=KL+\ol{H}+D^C+\sum_{|\{i,j\}\cap\{k\ell\}|=1} IJ,\] so $\mathcal{L}_{k\ell}$ is singular if and only if $KL+\ol{H}$ is singular and $D^C=\sum IJ=0$, as $KL+\ol{H}$ has eigenvalues in the closed left-hand plane (by the same argument as before) and $D^C$ and each $IJ$ is negative semidefinite. The same logic as before shows this happens if and only if $h_k=h_\ell$, $c_{nn}=0$ for all $k-1\leq n\leq \ell-1$, and either $k\ell\in S^2_{G_\mathcal{L}}$ or $k,\ell\in S_{G_\mathcal{L}}$, in which case \[\ker\mathcal{L}_{k\ell}=\ker KL=\left\{
\begin{array}{ll}
\C\{(c_{k\ell}+a_{k\ell})\lambda_{k\ell}+(c_{\ell k} +a_{k\ell})\lambda_{\ell k}\} & \textrm{if } k\ell\in S^2_{G_\mathcal{L}}\\
\Span(\lambda_{k\ell},\lambda_{\ell k}) & \textrm{if } k,\ell\in S_{G_\mathcal{L}} \\
\end{array} 
\right..\]
\end{proof}

The next two statements follow similarly to Corollary~\ref{negative E} and Theorem~\ref{eigenvalues E}.

\begin{corollary} \label{diag C gives hermitian}
	Let $\mathcal{L}$ be a Hamiltonian-free QDS generator which is pair block diagonal with respect to the Gell-Mann basis \eqref{GM form} with $C^{\mathcal{D}_0}$ diagonal. Then $\mathcal{L}^\mathcal{O}$ is negative semidefinite.
\end{corollary}

\begin{remark} \label{eigenvalues C} If $\mathcal{L}$ is a QDS generator which is pair block diagonal with respect to the Gell-Mann basis \eqref{GM form} with $H=\sum_{n=1}^Nh_nE_{nn}$ and $C^{\mathcal{D}_0}$ diagonal, then the $k\ell$ block $\mathcal{L}_{k\ell}$ of $\mathcal{L}$ has eigenmatrices \begin{align*}A^{\pm} = &\left[a_{k\ell}+\frac{c_{k\ell}-c_{\ell k}}{2}-(h_k-h_\ell)\pm\sqrt{\left(\frac{c_{k\ell}-c_{\ell k}}{2}\right)^2+a_{k\ell}^2-(h_k-h_\ell)^2}\right]\lambda_{k\ell} \\ + &\left[a_{k\ell}-\frac{c_{k\ell}-c_{\ell k}}{2}+(h_k-h_\ell)\pm\sqrt{\left(\frac{c_{k\ell}-c_{\ell k}}{2}\right)^2+a_{k\ell}^2-(h_k-h_\ell)^2}\right]\lambda_{\ell k}\end{align*}corresponding to eigenvalues $\mu^{\pm}=$\begin{align*}&-\frac{1}{2}\left(c_{k\ell}+c_{\ell k}+\frac{k-1}{k}c_{k-1,k-1}+\sum_{m=k}^{\ell-2}\frac{1}{{m(m+1)}}c_{mm}+\frac{\ell}{\ell-1}c_{\ell-1,\ell-1}\right.\\ &\left.+\frac{1}{2}\!\!\sum_{i\not\in\{k,\ell\}\ni j}\!\!\!\!(c_{ij}+c_{ji}- 2b_{ij})+\frac{1}{2}\!\!\sum_{i\in\{k,\ell\}\not\ni j}\!\!\!\!(c_{ij}++c_{ji}+ 2b_{ij})\right)\pm \sqrt{\left(\frac{c_{k\ell}-c_{\ell k}}{2}\right)^2+a_{k\ell}^2-(h_k-h_\ell)^2}.\end{align*} In particular, both of $\lambda_{k\ell}$ and $\lambda_{\ell k}$ are eigenmatrices of $\mathcal{L}^{\mathcal{O}}$ if and only if $h_k-h_\ell=a_{k\ell}=0$, in which case they have eigenvalues $-c_{\ell k}-\mu$ and $-c_{k \ell}-\mu$, respectively, where $2\mu=$ \footnotesize\begin{align*}\frac{1}{2}\!\!\sum_{i\not\in\{k,\ell\}\ni j}\!\!\!\!(c_{ij}+c_{ji}- 2b_{ij})+\frac{1}{2}\!\!\sum_{i\in\{k,\ell\}\not\ni j}\!\!\!\!(c_{ij}+c_{ji}+ 2b_{ij})+\frac{k-1}{k}c_{k-1,k-1}+\sum_{m=k}^{\ell-2}\frac{1}{{m(m+1)}}c_{mm}+\frac{\ell}{\ell-1}c_{\ell-1,\ell-1}\!.\end{align*} \end{remark}

One might compare this last remark to Theorem~5 of \cite{S}, where Siudzi\'nska determines the eigenvalues of a QDS generator $\mathcal{L}$ which is written in Gell-Mann form \eqref{GM form} with $H=0$ and $C$ diagonal, and for which every $\lambda_{ij}$ (including $i=j$) is an eigenmatrix of $\mathcal{L}$.

In the case $C^\mathcal{O}$ is diagonal the digraph $G_\mathcal{L}$ satisfies $\gamma_{ij}=\gamma_{ji}$ for all vertices $i$ and $j$, and hence $G_\mathcal{L}$ may be regarded as an (undirected) graph $H_\mathcal{L}$. Let $I_{H_\mathcal{L}}$ denote the set of isolated vertices of $H_\mathcal{L}$, and let $I^2_{H_\mathcal{L}}$ denote the set of isolated edges $k\ell$ of $H_\mathcal{L}$ for which $c_{k\ell}c_{\ell k}=0$ (i.e., the set singular 2-sinks ignoring direction). The statement of Lemma~\ref{diag LSA} is simplified to the following:
 
\begin{corollary} \label{diagg LSA}
	Let $\mathcal{L}$ be a QDS generator written with respect to the Gell-Mann basis \eqref{GM form} such that $H=\sum_{n=1}^N h_n E_{nn}$ and $C$ is diagonal. Then the $k\ell$ block $\mathcal{L}_{k\ell}$ of $\mathcal{L}$ is singular if and only if $h_k=h_\ell$, $c_{nn}=0$ for $k-1\leq n\leq \ell-1$, and \begin{itemize}
		\item $k,\ell\in I_{H_\mathcal{L}}$, in which case $\ker\mathcal{L}_{k\ell}=\Span(\lambda_{k\ell},\lambda_{\ell k})$, or
		\item $k\ell\in I^2_{H_\mathcal{L}}$, in which case \begin{itemize}
			\item $\ker\mathcal{L}_{k\ell}=\C\{\lambda_{k\ell}\}$ if $c_{\ell k}=0$,
			\item $\ker\mathcal{L}_{k\ell}=\C\{\lambda_{\ell k}\}$ if $c_{k\ell }=0$.
	\end{itemize}	\end{itemize}
\end{corollary}

\subsection{Examining the Full Generator $\boldsymbol{\mathcal{L}}$}\label{section L}

To establish the final kernel results for this section, we need only recall that pair block diagonal generators are of form~\eqref{block L}. From Corollary~\ref{general ker LDI} and Lemma~\ref{LSA in E}, we have the following:

\begin{theorem} \label{main pair block in E}
	Let $\mathcal{L}$ be a QDS generator which is pair block diagonal with respect to the standard basis \eqref{E form} with $\widetilde{H}=\sum_{n=1}^Nh_nE_{nn}$ and $\Gamma\geq 0$. Then \[\ker\mathcal{L}=\bigoplus_{k,\ell}\ker\mathcal{L}_{k\ell}\oplus\Span\left(d^{G_\mathcal{L}^n}\right)_{n=1}^{k},\] where $d^{G_\mathcal{L}^n}$ are given by \eqref{d_n} and $\ker\mathcal{L}_{k\ell}$ are as in Lemma~\ref{LSA in E}. 
\end{theorem}

\begin{theorem} \label{main SA block} Let $\mathcal{L}$ be a QDS generator which is pair block diagonal with respect to the Gell-Mann basis \eqref{GM form} with $H=\sum_{n=1}^Nh_nE_{nn}$ and $C^{\mathcal{D}_0}$ diagonal. Then \[\ker\mathcal{L}=\bigoplus_{k,\ell}\ker\mathcal{L}_{k\ell}\oplus\Span\left(d^{G_\mathcal{L}^n}\right)_{n=1}^{k},\] where $d^{G_\mathcal{L}^n}$ are given by \eqref{d_n} and $\ker\mathcal{L}_{k\ell}$ are as in Lemma~\ref{diag LSA}.\end{theorem}

\begin{corollary} \label{main diag} Let $\mathcal{L}$ be a QDS generator written with respect to the Gell-Mann basis \eqref{GM form} such that $H=\sum_{n=1}^N h_n E_{nn}$ and $C$ is diagonal. Then \[\ker\mathcal{L}=\bigoplus_{k,\ell}\ker\mathcal{L}_{k\ell}\oplus\Span\left(d^{H_\mathcal{L}^n}\right)_{n=1}^{k},\] where $d^{H_\mathcal{L}^n}$ are given by \eqref{tilded_n} and $\ker\mathcal{L}_{k\ell}$ are as in Corollary~\ref{diagg LSA}.
\end{corollary}

Recalling \eqref{kerstates}, these Theorems allow us to compute exactly the invariant states for pair block diagonal generators with diagonal Hamiltonian from statistics of the underlying graph. Namely, the diagonal entries are computed from the total weight of spanning trees rooted at each vertex, and the off-diagonal entries arise from the presence of sinks and singular 2-sinks. Examples~\ref{manifest} and \ref{superpositionexample} below illustrate how these various structures in the associated digraph $G_\mathcal{L}$ affect the structure of the invariant states.

\begin{example}\label{manifest}
	In dimension $N=8$, consider QDS generator $\mathcal{L}$ given by \eqref{E form} with Hamiltonian $H=\sum_{i=1}^8 h_iE_{ii}$ with $h_2=h_3$ and $h_4=h_5$, and coefficient matrix $\Gamma$ whose entries are all zero except the 45 block given by $\left(\begin{smallmatrix}
	1 & \imath \\ -\imath & 1
	\end{smallmatrix}\right)$ and the 67, 68, and 78 blocks given by $\left(\begin{smallmatrix}
	1 & 0\\ 0 & 2
	\end{smallmatrix}\right)$, $\left(\begin{smallmatrix}
	3 & 0\\ 0 & 3
	\end{smallmatrix}\right)$, and $\left(\begin{smallmatrix}
	4 & 0\\ 0 & 1
	\end{smallmatrix}\right)$ respectively. The graph $G_\mathcal{L}$ is drawn below, where the dashed edge is a singular 2-sink.
	
	 \begin{minipage}[t]{0.5\textwidth}\[\!\!\!\!\!\!\!\!\!\!\!\!\!\!\!\!\!\!\!\!\!\!\!\!
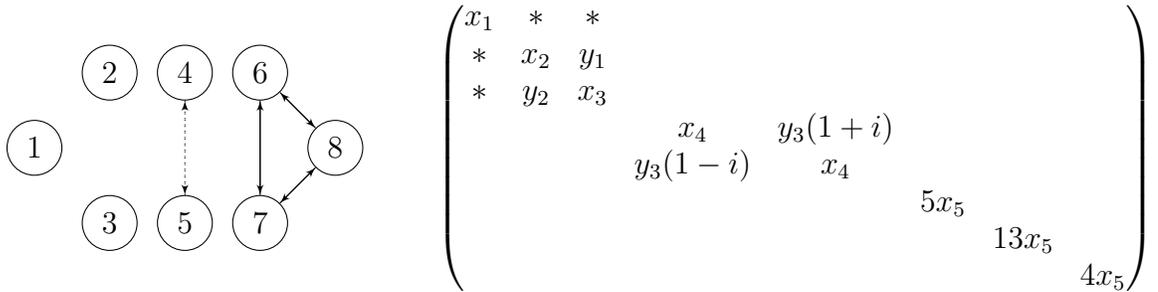
\begin{tikzpicture}
	\tikzset{vertex/.style = {shape=circle,draw,minimum size=1.5em}}
	\tikzset{edge/.style = {->,> = latex'}}
	
	\node[vertex] (1) at (1,2) {1};
	\node[vertex] (2) at (2,3) {2};
	\node[vertex] (3) at (2,1) {3};
	\node[vertex] (4) at (3,3) {4};
	\node[vertex] (5) at (3,1) {5};
	\node[vertex] (6) at (4,3) {6};
	\node[vertex] (7) at (4,1) {7};
	\node[vertex] (8) at (5,2) {8};
	
	\node (9) at (3,4) {};
	
	\foreach \from/\to in {4/5,5/4}
	\draw[edge,dotted] (\from) -> (\to);
	\foreach \from/\to in {6/7,7/6,6/8,8/6,7/8,8/7}
	\draw[edge] (\from) -> (\to);
	
	\end{tikzpicture}\]
\end{minipage}\begin{minipage}[t]{0.5\textwidth}\[\!\!\!\!\!\!\!\!\!\!\!\!\!\!\!\!\!\!\!\!\!\!\!\!\!\!\!\!\begin{pmatrix}
x_1 & \ast & \ast &  &  &  &  &  \\ 
\ast & x_2 & y_1 &  &  &  &  &  \\ 
\ast & y_2 & x_3 &  &  &  &  &  \\ 
&  &  & x_4 & y_3(1+i) &  &  &  \\ 
&  &  & y_3(1-i) & x_4 &  &  &  \\ 
&  &  &  &  & 5x_5 &  &  \\ 
&  &  &  &  &  & 13x_5 &  \\ 
&  &  &  &  &  &  & 4x_5
\end{pmatrix}\]
\end{minipage}

The kernel of $\mathcal{L}$ can be computed via Theorem~\ref{main pair block in E}, where each pair of $(k,\ell)$ and $(\ell,k)$ entries are given by $\ker\mathcal{L}_{k\ell}$. The displayed matrix represents an arbitrary element in $\ker\mathcal{L}$ where missing entries are zero. Specifically, the five $x_n$'s represent multiples of $d^{G_\mathcal{L}^n}$ for each of the five TSCCs, computed as in \eqref{d_n}, and the $y_n$'s represent multiples of the off-diagonal kernel elements described in Lemma~\ref{LSA in E}. The entries denoted by $\ast$ represent zero if $h_1\neq h_2,h_3$, or additional free variables if $h_1=h_2=h_3$. Notice that one may create both non-faithful and/or non-diagonal invariant states. Notice also that the presence of a singular 2-sink puts relations on the real and imaginary parts of certain off-diagonal coordinates of the kernel elements, a phenomenon that does not happen in the graph induced case \eqref{generic}.
\end{example}

\begin{example}\label{superpositionexample} Consider a system with three states: $|1\rangle$, $|2\rangle$, and $|3\rangle$. Consider the jump between $\frac{1}{\sqrt{2}}(|1\rangle+\imath|2\rangle)\mapsto\frac{1}{\sqrt{2}}(\imath|1\rangle+|2\rangle)$ at rate $a>0$ together with the jumps $|3\rangle\mapsto|1\rangle$ at rate $b>0$ and $|3\rangle\mapsto|2\rangle$ at rate $c>0$. Following Remark~\ref{superposition}, we model this by setting the entries of coefficient matrix $\Gamma$ all zero except the 12 block given by $\left(\begin{smallmatrix}
	a & a \\ a & a
	\end{smallmatrix}\right)$, 13 block given by $\left(\begin{smallmatrix}
	b & 0\\ 0 & 0
	\end{smallmatrix}\right)$, and the 23 block given by $\left(\begin{smallmatrix}
	c & 0\\ 0 & 0
	\end{smallmatrix}\right)$. Applying Theorem~\ref{main pair block in E}, we have that $$\ker\mathcal{L}=\Span\left(\left(\begin{smallmatrix}
	1 & 0 & 0 \\ 0 & 1 & 0 \\ 0 & 0 & 0
	\end{smallmatrix}\right),\left(\begin{smallmatrix}
	0 & 1 & 0 \\ 1 & 0 & 0 \\ 0 & 0 & 0
	\end{smallmatrix}\right)\right),$$ and so the invariant states of this system are given by $$\frac{1}{2}\begin{pmatrix}
	1 & x & 0 \\ 
	x & 1 & 0 \\ 
	0 & 0 & 0
	\end{pmatrix}$$ for any $-1\leq x\leq 1$. In particular, $\frac{1}{\sqrt{2}}(|1\rangle+|2\rangle)$ is an invariant state. In the graph induced case \eqref{generic}, i.e. if only jumps between vector states $|i\rangle\mapsto|j\rangle$ had been allowed, this could only happen in the trivial case that the jump rates for $|1\rangle\mapsto|i\rangle$ and $|2\rangle\mapsto|i\rangle$ were identically zero for all $i$. Allowing jumps between superpositions thus enables the system to maintain coherence despite nontrivial evolution.
\end{example}

\section{Other Generators} \subsection{Identity Preserving QDSs}\label{maxdisssection}

In this section we examine QDSs whose generators satisfy $\mathcal{L}(I_N)=0$; that is, QDSs for which the maximally mixed state $I_N/N$ is invariant, or, equivalently by Corollary~\ref{diss}, QDSs which are contractive for some/all $p$-Schatten norm with $p>1$. We prove that the kernel of such a QDS generator is contained in the kernel of a second, naturally induced QDS generator which is characterized by Corollary~\ref{main diag}. To define this second generator we first consider the kernel of the coefficient matrix $C$ for $\mathcal{L}$ written in Gell-Mann form \eqref{GM form}.

\begin{lemma} \label{subtract e} Let $C:M_N^0(\C)\to M_N^0(\C)$ with $C\geq0$, and let $x_1,\ldots, x_n\in M_N^0(\C)$ be orthonormal in $S_2^N$. Then $C-\epsilon\sum_{i=1}^n |x_i \rangle\langle x_i|\geq0$ for some $\epsilon>0$ if and only if $\{x_1,\ldots,x_n\}\subseteq  (\ker C)^\perp$.
\end{lemma}

\begin{proof} 
	Let $\epsilon=\inf_{y\in (\ker C)^\perp, ||y||=1}\langle y,Cy\rangle$. That $\epsilon \geq 0$ is clear since $C\geq 0$. We claim that $\epsilon>0$. Indeed, the unit ball of $(\ker C)^\perp$ is compact (being finite dimensional) and so the infimum is achieved at some $y_0\in (\ker C)^\perp$. Since $Cy_0\neq 0$ we have $\sqrt{C}y_0\neq0$, and hence $\langle y_0, Cy_0\rangle=\langle \sqrt{C}y_0, \sqrt{C}y_0\rangle=||\sqrt{C}y_0||^2\neq 0$.
	
	Now, suppose $\{x_1,\ldots,x_n\}$ is an orthonormal subset of $(\ker C)^\perp$ and let $\{k_1\ldots,k_m\}$ be an orthonormal basis of $\ker C$. Then there exist $x_{n+1},\ldots, x_{\ell}\in M_N^0(\C)$ such that $\{k_1,\ldots, k_m,x_1,\ldots,x_\ell\}$ is an orthonormal basis of $M_N(\C)$. Letting $z\in M_N^0(\C)$ we aim to show $\langle z, (C-\epsilon\sum_{i=1}^n|x_i\rangle\langle x_i|)z\rangle\geq 0$. Indeed, writing $z=\sum_{s=1}^m a_s k_s+\sum_{t=1}^\ell b_t x_t$ we may define $\widetilde z:=\sum_{t=1}^\ell b_t x_t$ and assume $||\widetilde z||^2=\sum_{t=1}^\ell |b_t|^2=1$ without loss of generality. Then $C=C^\ast$ and $Cz=C\widetilde z$ imply $$\langle z,Cz\rangle =\langle z,C\widetilde z\rangle =\langle C z,\widetilde z\rangle=\langle C\widetilde z,\widetilde z\rangle =\langle \widetilde z,C\widetilde z\rangle\geq \epsilon,$$ and so \begin{align*} \langle z, (C-\epsilon\sum_{i=1}^n|x_i\rangle\langle x_i|)z\rangle &= \langle z, Cz\rangle -\epsilon \sum_{i=1}^n \langle z, |x_i\rangle\langle x_i|z\rangle \\& = \langle z, Cz\rangle -\epsilon \sum_{t=1}^\ell |b_t|^2 = \langle z, Cz\rangle -\epsilon\geq 0.
	\end{align*}
	
	Conversely, suppose $\{x_1,\ldots,x_n\}\not\subseteq(\ker C)^\perp$ so there is some $k\in \ker C$ such that $k\not\perp x_j$ for some $1\leq j\leq n$. Then $|\langle k,x_j\rangle|^2>0$, and so for all $\epsilon>0$ we have \begin{align*} \langle k, (C-\epsilon\sum_{i=1}^n|x_i\rangle\langle x_i|)k\rangle = \langle k, Ck\rangle -\epsilon \sum_{i=1}^n \langle k, |x_i\rangle\langle x_i|k\rangle = -\epsilon \sum_{i=1}^n |\langle k,x_i\rangle|^2 <0.\end{align*}
\end{proof}

\begin{remark} \label{K} Let $\mathcal{L}$ be a QDS generator written in Gell-Mann form \eqref{GM form} with coefficient matrix $C$, and define $K:M_N^0\to M_N^0$ by $K=\sum|\lambda_{ij} \rangle\langle \lambda_{ij}|$, where the sum is over all $\lambda_{ij}$ perpendicular to $\ker C$. Then $C-\epsilon K\geq0$ for some $\epsilon>0$. Further, $K\geq0$ and so taking $K$ to be the coefficient matrix in Gell-Mann form \eqref{GM form} defines a QDS generator $\mathcal{K}$ by Theorem~\ref{GKSL}. Since $K$ is diagonal we have $\mathcal{K}$ is of form \eqref{block L2}, $\mathcal{K}(I_N)=0$, and further $\mathcal{K}$ is negative semidefinite by Remark~\ref{LD hermitian} and Corollary~\ref{diag C gives hermitian}. \end{remark}

\begin{proposition} \label{kerL<kerK} Let $\mathcal{L}$ be a QDS generator satisfying $\mathcal{L}(I_N)=0$. Then \[\ker\mathcal{L}\subseteq \ker\mathcal{K},\] where $\ker\mathcal{K}$ is given by Corollary~\ref{main diag}.
\end{proposition}

\begin{proof} Fix $\epsilon>0$ such that $C-\epsilon K\geq 0$. It is easy to see that using $C-\epsilon K$ as the coefficient matrix in Gell-Mann form \eqref{GM form} gives rise to the QDS generator $\mathcal{L}-\epsilon \mathcal{K}$, and that $\mathcal{L}=(\mathcal{L}-\epsilon \mathcal{K})+\mathcal{K}$. The result then follows from Lemma~\ref{L=A+B}.
\end{proof}

We note that $\mathcal{L}$ does not need to be written in Gell-Mann form \eqref{GM form} to define $\mathcal{K}$, as our definition relies only on the kernel of the coefficient matrix $C$. Recalling that Theorem~\ref{GKSL} uniquely defines $C$ (as an operator), or more generally that Theorem~\ref{GKSL3} uniquely defines $\Gamma$, this kernel is uniquely defined regardless of basis $\{F_i\}$.

\subsection{Consistent Generators.} \label{consistentsection}

In this section we examine those generators for which the Hamiltonian $H$ is `well-behaved'. More precisely, let $H_\mathcal{L}$ denote the graph obtained from $G_\mathcal{L}$ by ignoring weights and directedness of the edges, and for each connected component $H_\mathcal{L}^k$ of $H_\mathcal{L}$ let $P_k$ be the orthogonal projection onto $\Span(E_{ij})_{i,j\in V(H_\mathcal{L}^k)}$. We call $H$ {\bf consistent} if $P_kHP_\ell=0$ for all $\ell\neq k$. We provide a lower bound for the dimension of the kernel of a QDS generator for which $H$ is consistent.
	
Recall that the definition of a QDS immediately implies $\Tr(\mathcal{L}(A))=0$ for all $A\in M_N(\C)$. The next result says that certain submatrices of $\mathcal{L}(A)$ are also traceless if we assume the Hamiltonian $H$ is consistent.

\begin{theorem} \label{proj}
	Let $\mathcal{L}$ be a QDS generator. Considering fixed $k$, if $P_kHP_\ell=0$ for all $\ell\neq k$, then $\Tr(P_k\mathcal{L}(A))=0$ for all $A\in M_N(\C)$.
\end{theorem}

\begin{proof}
	Consider $\mathcal{L}$ written with respect to the standard basis \eqref{E form} such that $\Gamma$ satisfies the conditions of Theorem~\ref{GKSL3}. If $H_\mathcal{L}$ is connected then the statement is obvious since $\mathcal{L}$ has traceless range, so assume that $H_\mathcal{L}$ is not connected and $H_\mathcal{L}^n$, $H_\mathcal{L}^m$ are distinct connected components. Then for any $i\in V(H_\mathcal{L}^n)$ and $j\in V(H_\mathcal{L}^m)$ we have that weights $\gamma_{ij}=\gamma_{ji}=0$. Further, positive semidefiniteness of $\Gamma$ implies that each entry of $\Gamma$ which shares a row or column with $\gamma_{ijij}$ or $\gamma_{jiji}$ is also zero (for if not the $2\times 2$ submatrix formed by removing all other rows and columns would have negative determinant, contradicting positive semidefiniteness). Hence \begin{equation}\label{consistent}\mathcal{L}=-\imath[H,\cdot]+\sum_{n,m}\underset{k,\ell\in V(H_\mathcal{L}^m)}{\sum_{i,j\in V(H_\mathcal{L}^n)}} \gamma_{ijk\ell}D_{ijk\ell}.\end{equation}
	By linearity of $\mathcal{L}$ it suffices to show $\Tr(P_k\mathcal{L}(E_{st}))=0$ for arbitrary $1\leq s, t \leq N$. To this end, we claim that every output $\mathcal{L}(E_{st})$ which has nonzero diagonal is traceless with its nonzero diagonal in $\Span(E_{nn})_{n\in V(H_\mathcal{L}^m)}$ for some $m$. Since each output of $\mathcal{L}$ is a linear combination of outputs of  $[H,\cdot]$ and of the $D_{ijk\ell}$ appearing in \eqref{consistent}, it suffices to show this for $[H,\cdot]$ and those $D_{ijk\ell}$ separately.
	
	For the Hamiltonian part we write $H=\sum h_{ij}E_{ij}$ so that $[H,\cdot]=\sum h_{ij}[E_{ij},\cdot]$. Note that if $P_kHP_\ell= 0$ for $k\neq \ell$ then for any $i\in V(H_\mathcal{L}^k)$ and $j\in V(H_\mathcal{L}^\ell)$ we have $h_{ij}=0$. That is, if $h_{ij}\neq 0$ then $i,j\in V(H_\mathcal{L}^m)$ for some $m$. From this the claim is clear, as $[E_{ij},E_{st}]$ has nonzero diagonal output if and only if $i=t$ and $j=s$, in which case $[E_{ij},E_{st}]=E_{ii}-E_{jj}$.
	
	For the operators $D_{ijk\ell}$ we recall \eqref{E calculation}, which reads \[
	\notag D_{ijk\ell}(E_{st}) = 2\delta_{js}\delta_{\ell t}E_{ik}-\delta_{ik}\delta_{js}E_{\ell t}-\delta_{\ell t}\delta_{ik}E_{sj}.
	\] Thus, $ D_{ijk\ell}(E_{st})$ has nonzero diagonal if and only if $i=k$, $j=s$, and $\ell=t$, in which case $D_{ijk\ell}(E_{st})=2E_{ii}-E_{jj}-E_{\ell\ell}$. If $D_{ijk\ell}$ appears in \eqref{consistent}, then these equalities imply $i,j,\ell\in V(H_\mathcal{L}^m)$ for some $m$.

\end{proof}

\begin{corollary} \label{cc<dimkerL} Let $\mathcal{L}$ be a QDS generator such that $H$ is consistent. Then \[cc(H_\mathcal{L})\leq\dim\ker\mathcal{L},\] where $cc(H_\mathcal{L})$ is the number of connected components of $H_\mathcal{L}$.
\end{corollary}

\begin{proof} Consider the connected components $H^1_\mathcal{L},\ldots,H^\ell_\mathcal{L}$ of $H_\mathcal{L}$ ordered so that $|V(H^n_\mathcal{L})|\geq2$ for $n\leq m$ and $|V(H^n_\mathcal{L})|=1$ for $n>m$ for some $m\geq 0$. It suffices to find $\ell$ many pairwise orthogonal matrices not in $\Range(\mathcal{L})$. Since $H$ is consistent, by Theorem~\ref{proj} we have $\Tr(P_k\mathcal{L}(A))=0$ for all $A\in M_N(\C)$ and all $1\leq k\leq \ell$. In the boundary case of $m=0$ we have that $E_{ii}\not\in\Range(\mathcal{L})$ for all $1\leq i\leq N$ and so $\ell=N\leq \dim\ker \mathcal{L}$. Otherwise, if $m>1$, fixing $i_1\in V(H_\mathcal{L}^1)$ and $j_2\in V(H_\mathcal{L}^2)$ we have $E_{i_1i_1}-E_{j_2j_2}\not\in\Range(\mathcal{L})$. Similarly, fixing some $i_2\in V(H_\mathcal{L}^2)\setminus\{j_2\}$ and $j_3\in V(H_\mathcal{L}^3)$ we have $E_{i_2i_2}-E_{j_3j_3}\not\in\Range(\mathcal{L})$. We continue until we find $E_{i_mi_m}-E_{j_{m+1}j_{m+1}}\not\in\Range(\mathcal{L})$, for a total of $m$ simple differences $E_{ii}-E_{jj}$ not in $\Range(\mathcal{L})$. Further, writing $V(H_\mathcal{L}^n)=\{i_n\}$ for all $n\geq m+2$ we have $E_{i_ni_n}\not\in\Range(\mathcal{L})$, for a total of $\ell-m-1$ distinct $E_{ii}$ not in $\Range(\mathcal{L})$. Because these chosen matrices are all diagonal and we have no repeated indices, we have a set of $\ell-1$ pairwise orthogonal matrices. It is clear that $I_N-\sum_{m+2\leq n\leq \ell}E_{i_ni_n}$ is nonzero and orthogonal to the above matrices, and is not in $\Range(\mathcal{L})$ since $\mathcal{L}$ has traceless range, and so we have found a set of $\ell$ many orthogonal matrices not in $\Range(\mathcal{L})$, as desired.
	\end{proof}

Since certainly a QDS is not uniquely relaxing if it has multiple invariant states, we immediately have the following.

\begin{corollary} \label{necc connected} Let $\mathcal{L}$ be a QDS generator such that $H$ is consistent. If $T_t$ is uniquely relaxing then $H_\mathcal{L}$ is connected.\end{corollary}

We note that it is not true that the number of TSCCs of $G_\mathcal{L}$ lower bounds $\dim\ker\mathcal{L}$ in general, even with consistent $H$; for example, see the example of section 2 of aforementioned \cite{Glos} for which $G_\mathcal{L}$ has two TSCCs yet the QDS has a single invariant state.

\section{Conclusion}

We began this work by determining when the famed GKSL form \eqref{GKSLeq} would define a QDS generator when allowed not necessarily traceless operators $F_i$ (Theorem~\ref{GKSL2}). Along the way, we identified that the coefficient matrix $C$ of the classical GKSL form \eqref{GKSLeq} is uniquely determined by $\mathcal{L}$ when viewed as an operator (discussion above Theorem~\ref{GKSL2}), but this is not necessarily true for the coefficient matrix $\Gamma$ of the more general form \eqref{GKSL2eq} unless stronger assumptions are met (Theorem~\ref{GKSL3}). In any case, these theorems offer criteria for when $\mathcal{L}$ written with respect to the standard basis \eqref{E form} defines a QDS generator, a form whose simplicity is advantageous for both calculation and understanding.

With this easy to work with form, we established the class of pair block diagonal generators (Definition~\ref{convention}) to generalize the graph induced generators given by \eqref{generic} while preserving the important properties, such as leaving the diagonal subalgebra $\mathcal{D}$ and off-diagonal subspace $\mathcal{O}$ invariant in the case of diagonal Hamiltonian $H$. We also established the synonymous definition in terms of the Gell-Mann basis (Definition~\ref{convention2}), which is often used due to its traceless construction when dealing with the GKSL form \eqref{GKSL}.

For the class of pair block diagonal generators, we found explicit formula for all invariant states when the Hamiltonian is diagonal (Theorem~\ref{main pair block in E}), and furthermore all eigenmatrices which belong to the off-diagonal subspace $\mathcal{O}$ and their corresponding eigenvalues (Theorem~\ref{eigenvalues E}). In particular, the invariant states depend on the structure of a naturally induced digraph. Though we do not explore such applications in this work, we note that these results allow for exact computation of rates of convergence of such QDSs, given the Laplacian spectral gap of the induced digraph.

We have also shown explicitly that, when written in matrix form, every QDS generator contains as a submatrix a naturally associated digraph Laplacian (Theorem~\ref{LD}). In the case the Hamiltonian is consistent with this digraph, connectedness properties of the digraph identify submatrices of elements in the range of $\mathcal{L}$ as traceless (Theorem~\ref{proj}), and hence we have established lower bounds on the number of invariant states of the QDS based on the connectedness properties of the digraph (Corollary~\ref{cc<dimkerL}). In the case the maximally mixed state is invariant, which happens if and only if the QDS is contraction in some/all $p$-Schatten norms with $p>1$ (Corollary~\ref{diss}), we have shown that the structure of the invariant states can be inferred from the digraph naturally associated to the kernel of the coefficient matrix (Proposition~\ref{kerL<kerK}).

\appendix
\section{Calculations for $D^\lambda_{nn}$}\label{Diijj appen} 

We use $\delta_{i\leq j}$ to denote the indicator \[\delta_{i\leq j}=\left\{\begin{array}{ll}
1 & \textrm{if } i\leq j \\
0 & \textrm{otherwise} \\
\end{array} 
\right.,\] and similarly for $\delta_{i\leq j\leq k}$. Since diagonal matrices commute we have $D_{nnmm}(\lambda_{\ell\ell})=0$ for all $n,m,\ell$. Thus, for $k<\ell$,
	
	\[ D^\lambda_{nn}(\lambda_{k\ell}) =[\lambda_{nn},\lambda_{k\ell}\lambda_{nn}]+[\lambda_{nn}\lambda_{k\ell},\lambda_{nn}]=2\lambda_{nn}\lambda_{k\ell}\lambda_{nn}-\lambda_{k\ell}\lambda_{nn}\lambda_{nn}-\lambda_{nn}\lambda_{nn}\lambda_{k\ell},\] where \begin{align*}2\lambda_{nn}\lambda_{k\ell}\lambda_{nn}	&= \textstyle\frac{2}{\sqrt{2}n(n+1)}\displaystyle\left(\sum_{m=1}^n E_{mm}-n E_{n+1,n+1}\right)(E_{k\ell}+E_{\ell k})\left(\sum_{m=1}^n E_{mm}-n E_{n+1,n+1}\right)\\
	&=\textstyle\frac{2}{\sqrt{2}n(n+1)}\displaystyle(\delta_{k\leq n}E_{k\ell}{+}\delta_{\ell\leq n}E_{\ell k}{-}n\delta_{k,n+1}E_{k\ell}{-}n\delta_{\ell,n+1}E_{\ell k})\left(\sum_{m=1}^n E_{mm}{-}nE_{n+1,n+1}\right)\\
	&=\textstyle\frac{2}{\sqrt{2}n(n+1)}\Big(\delta_{k\leq n}\delta_{\ell\leq n}E_{k\ell}+\delta_{\ell\leq n}\delta_{k\leq n}E_{\ell k}-n\delta_{k,n+1}\delta_{\ell\leq n}E_{k\ell}-n\delta_{\ell,n+1}\delta_{k\leq n}E_{\ell k}\\
	&\phantom{{}={}}-n\delta_{k\leq n}\delta_{\ell,n+1}E_{k\ell}-n\delta_{\ell\leq n}\delta_{k,n+1}E_{\ell k}+n^2\delta_{k,n+1}\delta_{\ell,n+1}E_{k\ell}+n^2\delta_{\ell,n+1}\delta_{k,n+1}E_{\ell k}\Big)\\
	&=\textstyle\frac{2}{\sqrt{2}n(n+1)}(\delta_{\ell\leq n}E_{k\ell}+\delta_{\ell\leq n}E_{\ell k}-n\delta_{\ell,n+1}E_{\ell k}-n\delta_{\ell,n+1}E_{k\ell})\qquad\text{ using that $k<\ell$} \\
	&=\textstyle\frac{2}{n(n+1)}(\delta_{\ell\leq n}\lambda_{k\ell}-n\delta_{\ell,n+1}\lambda_{k \ell})\\
	&=\textstyle\frac{2}{n(n+1)}(\delta_{\ell\leq n}-n\delta_{\ell,n+1})\lambda_{k\ell}\end{align*}
	
	and $\lambda_{k\ell}\lambda_{nn}\lambda_{nn}+\lambda_{nn}\lambda_{nn}\lambda_{k\ell}=$ \begin{align*} &=\textstyle\frac{1}{\sqrt{2}n(n+1)}\displaystyle\Big((E_{k\ell}+E_{\ell k})\left(\sum_{m=1}^n E_{mm}+n^2 E_{n+1,n+1}\right){+}\left(\sum_{m=1}^n E_{mm}+n^2 E_{n+1,n+1}\right)(E_{k\ell}+E_{\ell k})\Big)\\
	&= \textstyle\frac{1}{\sqrt{2}n(n+1)}\Big((\delta_{\ell\leq n}E_{k\ell}+\delta_{k\leq n}E_{\ell k}+n^2\delta_{\ell,n+1}E_{k\ell}+n^2\delta_{k,n+1}E_{\ell k})\\&\phantom{{}={}}+(\delta_{k\leq n}E_{k \ell}+\delta_{\ell\leq n}E_{\ell k}+n^2\delta_{k,n+1}E_{k\ell}+n^2\delta_{\ell,n+1}E_{\ell k})\Big)\\
	&=\textstyle\frac{1}{n(n+1)}(\delta_{\ell\leq n}\lambda_{k\ell}+\delta_{k\leq n}\lambda_{k\ell}+n^2\delta_{\ell,n+1}\lambda_{k\ell}+n^2\delta_{k,n+1}\lambda_{k\ell})\\
	&=\textstyle\frac{1}{n(n+1)}(\delta_{\ell\leq n}+\delta_{k\leq n}+n^2\delta_{\ell,n+1}+n^2\delta_{k,n+1})\lambda_{k\ell}.\end{align*}
	
	Thus, \begin{align*}
	D^\lambda_{nn}(\lambda_{k\ell})&=\textstyle\frac{1}{n(n+1)}\Big(2(\delta_{\ell\leq n}-n\delta_{\ell,n+1})-(\delta_{\ell\leq n}+\delta_{k\leq n}+n^2\delta_{\ell,n+1}+n^2\delta_{k,n+1})\Big)\lambda_{k\ell}\\
		&= \textstyle\frac{1}{n(n+1)}(-n^2\delta_{k,n+1}-\delta_{k\leq n\leq \ell-2}-(n+1)^2\delta_{\ell,n+1})\lambda_{k\ell}\\
		&= \left\{
		\begin{array}{ll}
		\frac{-n}{(n+1)}\lambda_{k\ell} & n=k-1 \\
		\frac{-1}{n(n+1)}\lambda_{k\ell} & k\leq n\leq\ell-2 \\
		\frac{-(n+1)}{n}\lambda_{k\ell} & n=\ell-1 \\
		0 & \textrm{otherwise}
		\end{array} .
		\right. \end{align*}
	
	Similarly, 	\begin{align*}  D^\lambda_{nn}(\lambda_{\ell k})	&= \textstyle\frac{1}{n(n+1)}(-n^2\delta_{k,n+1}-\delta_{k\leq n\leq \ell-2}-(n+1)^2\delta_{\ell,n+1})\lambda_{\ell k}\\
	&= \left\{
	\begin{array}{ll}
	\frac{-n}{(n+1)}\lambda_{\ell k} & n=k-1 \\
	\frac{-1}{n(n+1)}\lambda_{\ell k} & k\leq n\leq\ell-2 \\
	\frac{-(n+1)}{n}\lambda_{\ell k} & n=\ell-1 \\
	0 & \textrm{otherwise}
	\end{array} .
	\right. \end{align*}

\bibliography{references} 

\begin{thebibliography}{10}

\bibitem{A}
Robert Alicki.
\newblock Invitation to quantum dynamical semigroups.
\newblock In Piotr Garbaczewski and Robert Olkiewicz, editors, {\em Dynamics of
  Dissipation}, pages 239--264. Springer Berlin Heidelberg, 2002.

\bibitem{AL}
Robert Alicki and Karl Lendi.
\newblock {\em Quantum dynamical semigroups and applications}, volume 717 of
  {\em Lecture Notes in Physics}.
\newblock Springer, Berlin, second edition, 2007.

\bibitem{GKS}
Vittorio Gorini, Andrzej Kossakowski, and E.~C.~G. Sudarshan.
\newblock Completely positive dynamical semigroups of {$N$}-level systems.
\newblock {\em J. Mathematical Phys.}, 17(5):821--825, 1976.

\bibitem{L}
G\"oran Lindblad.
\newblock On the generators of quantum dynamical semigroups.
\newblock {\em Comm. Math. Phys.}, 48(2):119--130, 1976.

\bibitem{RR}
James~D. Whitfield, C\'esar~A. Rodr\'{\i}guez-Rosario, and Al\'an Aspuru-Guzik.
\newblock Quantum stochastic walks: A generalization of classical random walks
  and quantum walks.
\newblock {\em Phys. Rev. A}, 81:022323, 2010.

\bibitem{LB}
Chaobin Liu and Radhakrishnan Balu.
\newblock Steady states of continuous-time open quantum walks.
\newblock {\em Quantum Inf. Process.}, 16(7):Art. 173, 11, 2017.

\bibitem{Pel}
Cl\'{e}ment Pellegrini.
\newblock Continuous time open quantum random walks and non-{M}arkovian
  {L}indblad master equations.
\newblock {\em J. Stat. Phys.}, 154(3):838--865, 2014.

\bibitem{SP}
Ilya Sinayskiy and Francesco Petruccione.
\newblock Microscopic derivation of open quantum walks.
\newblock {\em Phys. Rev. A (3)}, 92(3):032105, 11, 2015.

\bibitem{Glos}
Adam Glos, Jaros\l aw~Adam Miszczak, and Mateusz Ostaszewski.
\newblock Limiting properties of stochastic quantum walks on directed graphs.
\newblock {\em J. Phys. A}, 51(3):035304, 16, 2018.

\bibitem{AK}
Luigi Accardi and Sergei Kozyrev.
\newblock Lectures on quantum interacting particle systems.
\newblock In {\em Quantum interacting particle systems ({T}rento, 2000)},
  volume~14 of {\em QP--PQ: Quantum Probab. White Noise Anal.}, pages 1--195.
  World Sci. Publ., River Edge, NJ, 2002.

\bibitem{AFH}
Luigi Accardi, Franco Fagnola, and Skander Hachicha.
\newblock Generic {$q$}-{M}arkov semigroups and speed of convergence of
  {$q$}-algorithms.
\newblock {\em Infin. Dimens. Anal. Quantum Probab. Relat. Top.},
  9(4):567--594, 2006.

\bibitem{AKP}
Luigi Accardi, Franco Fagnola, and Skander Hachicha.
\newblock Generic {$q$}-{M}arkov semigroups and speed of convergence of
  {$q$}-algorithms.
\newblock {\em Infin. Dimens. Anal. Quantum Probab. Relat. Top.},
  9(4):567--594, 2006.

\bibitem{CSU}
Raffaella Carbone, Emanuela Sasso, and Veronica Umanit\`a.
\newblock Structure of generic quantum {M}arkov semigroup.
\newblock {\em Infin. Dimens. Anal. Quantum Probab. Relat. Top.},
  20(2):1750012, 19, 2017.

\bibitem{Evans}
David~E. Evans.
\newblock Irreducible quantum dynamical semigroups.
\newblock {\em Comm. Math. Phys.}, 54(3):293--297, 1977.

\bibitem{Dhahri}
Ameur Dhahri, Franco Fagnola, and Rolando Rebolledo.
\newblock The decoherence-free subalgebra of a quantum {M}arkov semigroup with
  unbounded generator.
\newblock {\em Infin. Dimens. Anal. Quantum Probab. Relat. Top.},
  13(3):413--433, 2010.

\bibitem{Carbone}
Raffaella Carbone, Emanuela Sasso, and Veronica Umanit\`a.
\newblock On the asymptotic behavior of generic quantum {M}arkov semigroups.
\newblock {\em Infin. Dimens. Anal. Quantum Probab. Relat. Top.},
  17(1):1450001, 18, 2014.

\bibitem{Carbone2}
Raffaella Carbone, Emanuela Sasso, and Veronica Umanit\`a.
\newblock Decoherence for positive semigroups on {$M_2(\Bbb C)$}.
\newblock {\em J. Math. Phys.}, 52(3):032202, 17, 2011.

\bibitem{BN}
Bernhard Baumgartner and Heide Narnhofer.
\newblock Analysis of quantum semigroups with {GKS}-{L}indblad generators.
  {II}. {G}eneral.
\newblock {\em J. Phys. A}, 41(39):395303, 26, 2008.

\bibitem{Garttner}
Martin G\"arttner.
\newblock Many-body effects in {R}ydberg gases: Coherent dynamics of strongly
  interacting two-level atoms and nonlinear optical response of a {R}ydberg gas
  in {EIT} configuration.
\newblock {\em PhD Thesis, Universit\"at Library Heidelberg}, 2013.

\bibitem{Pritchard}
Jonathan Pritchard.
\newblock {\em Cooperative Optical Non-Linearity in a Blockaded {R}ydberg
  Ensemble}.
\newblock Springer-Verlag Berlin Heidelberg, 2012.

\bibitem{Hofmann}
Christoph Hofmann.
\newblock Emergence of correlations in strongly interacting ultracold {R}ydberg
  gases.
\newblock {\em PhD Thesis, Universit\"at Library Heidelberg}, 2013.

\bibitem{P}
Vern Paulsen.
\newblock {\em Completely bounded maps and operator algebras}, volume~78 of
  {\em Cambridge Studies in Advanced Mathematics}.
\newblock Cambridge University Press, Cambridge, 2002.

\bibitem{EN}
Klaus-Jochen Engel and Rainer Nagel.
\newblock {\em One-parameter semigroups for linear evolution equations}, volume
  194 of {\em Graduate Texts in Mathematics}.
\newblock Springer-Verlag, New York, 2000.
\newblock With contributions by S. Brendle, M. Campiti, T. Hahn, G. Metafune,
  G. Nickel, D. Pallara, C. Perazzoli, A. Rhandi, S. Romanelli and R.
  Schnaubelt.

\bibitem{PG}
David P\'{e}rez-Garc\'{i}a, Michael~M. Wolf, Denes Petz, and Mary~Beth Ruskai.
\newblock Contractivity of positive and trace-preserving maps under {$L_p$}
  norms.
\newblock {\em J. Math. Phys.}, 47(8):083506, 5, 2006.

\bibitem{Lidar}
Daniel Lidar.
\newblock Lecture notes on the theory of open quantum systems, 02 2019.

\bibitem{H}
Timothy~F. Havel.
\newblock Robust procedures for converting among {L}indblad, {K}raus and matrix
  representations of quantum dynamical semigroups.
\newblock {\em J. Math. Phys.}, 44(2):534--557, 2003.

\bibitem{BK}
Reinhold~A. Bertlmann and Philipp Krammer.
\newblock Bloch vectors for qudits.
\newblock {\em J. Phys. A}, 41(23):235303, 21, 2008.

\bibitem{chung}
Fan R.~K. Chung.
\newblock {\em Spectral graph theory}, volume~92 of {\em CBMS Regional
  Conference Series in Mathematics}.
\newblock Published for the Conference Board of the Mathematical Sciences,
  Washington, DC; by the American Mathematical Society, Providence, RI, 1997.

\bibitem{tutte}
William~T. Tutte.
\newblock The dissection of equilateral triangles into equilateral triangles.
\newblock {\em Proc. Cambridge Philos. Soc.}, 44:463--482, 1948.

\bibitem{MG}
Inomzhon Mirzaev and Jeremy Gunawardena.
\newblock Laplacian dynamics on general graphs.
\newblock {\em Bull. Math. Biol.}, 75(11):2118--2149, 2013.

\bibitem{wu}
Chai~Wah Wu.
\newblock Algebraic connectivity of directed graphs.
\newblock {\em Linear and Multilinear Algebra}, 53(3):203--223, 2005.

\bibitem{S}
Katarzyna Siudzi\'{n}ska.
\newblock Two definitions of the {G}ell-{M}ann channels -- a comparative
  analysis.
\newblock {\em Rep. Math. Phys.}, 81(3):321--345, 2018.

\end{thebibliography}
\bibliographystyle{unsrt}

\end{document}